\newcommand*\patchAmsMathEnvironmentForLineno[1]{%
  \expandafter\let\csname old#1\expandafter\endcsname\csname #1\endcsname
  \expandafter\let\csname oldend#1\expandafter\endcsname\csname end#1\endcsname
  \renewenvironment{#1}%
     {\linenomath\csname old#1\endcsname}%
     {\csname oldend#1\endcsname\endlinenomath}}%
\newcommand*\patchBothAmsMathEnvironmentsForLineno[1]{%
  \patchAmsMathEnvironmentForLineno{#1}%
  \patchAmsMathEnvironmentForLineno{#1*}}%
\AtBeginDocument{%
\patchBothAmsMathEnvironmentsForLineno{equation}%
\patchBothAmsMathEnvironmentsForLineno{align}%
}
\patchAmsMathEnvironmentForLineno{equation}

\documentclass[aps,showpacs,superscriptaddress,floatfix,nofootinbib,twocolumn,longbibliography]{revtex4-1}

\usepackage[mathlines]{lineno}
\usepackage{amsmath,graphicx,verbatim,epsfig,amssymb,dsfont}
\usepackage{epstopdf}
\usepackage[latin1]{inputenc}
\usepackage{datetime}
\usepackage[colorlinks=true,linkcolor=blue,citecolor=blue]{hyperref}
\usepackage[usenames,dvipsnames]{color}
\usepackage{setspace}
\usepackage[mathscr]{eucal}

\usepackage{bbm}

\usepackage{float} 
\usepackage[caption=false]{subfig} 

\newcommand{\be}{\begin{equation}}
\newcommand{\bea}{\begin{eqnarray}}
\newcommand{\eea}{\end{eqnarray}}
\newcommand{\ee}{\end{equation}}

\usepackage[normalem]{ulem} 

\begin{document}

\title{Role of anomalous symmetry in $0$-$\pi$ qubits}

\author{I.L. Egusquiza}
\email{inigo.egusquiza@ehu.es}
\affiliation{Department of Physics, University of the Basque Country UPV/EHU, Apartado 644, 48080 Bilbao, Spain}
\affiliation{EHU Quantum Center, University of the Basque Country, UPV/EHU, Barrio Sarriena s/n, 48940 Leioa, Biscay, Spain}

\author{A. I\~niguez} 
\email{ainhoa.iniguez@ehu.eus}
\affiliation{Department of Mathematics, University of the Basque Country UPV/EHU, Apartado 644, 48080 Bilbao, Spain}

\author{E. Rico} 
\email{enrique.rico.ortega@gmail.com}
\affiliation{EHU Quantum Center, University of the Basque Country, UPV/EHU, Barrio Sarriena s/n, 48940 Leioa, Biscay, Spain}
\affiliation{Department of Physical Chemistry, University of the Basque Country UPV/EHU, Apartado 644, 48080 Bilbao, Spain}
\affiliation{IKERBASQUE, Basque Foundation for Science, Plaza Euskadi 5, 48009 Bilbao, Spain}

\author{A. Villarino}
\email{albavillarinopelaez@gmail.com}
\affiliation{Department of Physical Chemistry, University of the Basque Country UPV/EHU, Apartado 644, 48080 Bilbao, Spain}

\date{\today}

\begin{abstract}
We present an \emph{exact full symmetry analysis} of the $0$-$\pi$ superconducting circuit. We identify points in control parameter space of enhanced anomalous symmetry, which imposes robust twofold degeneracy of its ground-state, that is for \emph{all} values of the energy parameters of the model.  We show, both analytically and numerically, how this anomalous symmetry is maintained in the low-energy sector, thus providing us with a strong candidate for robust qubit engineering.
\end{abstract}

\maketitle 

\section{Introduction}

Symmetry arguments are the main organisational principle in our understanding of physical systems. In fact, not just the symmetries but their quantum representations set severe constraints on the theories that purport to describe our physical world \cite{Hooft1980,Harvey2005} as well as on their implementation on the lattice \cite{LesHouches2009}. Symmetry also allows us to characterise different phases of matter. Along these lines, and with insights from entanglement theory, symmetry-protected topological phases have been classified by the anomalous realisation of the symmetry at the spatial boundaries \cite{Fidkowski2011,Chen2012,Chen2013,Pollmann2012,Schuch2011}.

Recently, it has been discovered \cite{Gaiotto2017,Tanizaki2017} that also non-trivial examples with an anomalous representation of the symmetry exist in one-particle quantum mechanics. Following these works, we will characterise the symmetry properties of a self-protected superconducting circuit, the $0$-$\pi$ qubit \cite{Kitaev2006,Brooks2013}, and point at the anomalous realisation of the symmetry as a regime of interest to create symmetry-enhanced qubits, robust with respect to (at least some) design parameters.

The basic idea of the $0$-$\pi$ qubit is to search for a balancing so that only tunnelling of \emph{pairs} of Cooper pairs is possible. In the original proposals \cite{Doucot2002,Ioffe2002,Kitaev2006,Brooks2013}, it was realised that such a mechanism induces, in a particular design regime, two nearly degenerate ground-states, where the splitting of this degeneracy is exponentially small as a function of extensive system parameters, and stable with respect to weak local perturbations, which makes the $0$-$\pi$ qubit highly resistant to decoherence arising from local noise. Dempster et al. \cite{Dempster2014} provide a more detailed analysis of the relevant circuital proposal where they do have a hint of a symmetry analysis and Smith et al. \cite{Smith2020} present a semiclassical analysis. Also, in the last three years a series of works \cite{Groszkowski2018,Paolo2019} have led to the first experimental realisation of the $0$-$\pi$ qubit \cite{Gyenis2021}.

In this manuscript, we describe and understand the rich physics and symmetry properties of this superconducting qubit. At a point of special enhanced symmetry, the full spectrum is two-fold degenerate, and this feature is only dependent on the controlling offset charge and external flux, and completely independent of all the energy scales of the model. We first analyse a simpler system, the particle on the ring, which actually already presents the crucial aspects of this symmetry. We then study the $0$-$\pi$ circuit in this context, and show the connection in the low-energy sector.

\section{Discrete symmetries and anomaly}

As we shall see presently, all the essential symmetry properties of the $0$-$\pi$ qubit are realised in the $\cos(2\theta)$ qubit and its generalisations. Abstracting the corresponding Hamiltonians, let us consider one quantum particle moving on a ring, parametrised with $\theta\in[0,2\pi)$, under the action of a potential. In what follows, all operations with angles are to be understood modulo $2\pi$. As is well known, for any given potential there is a continuous $U(2)$ family of inequivalent Hamiltonians that all describe that particle \cite{Bonneau2001,Pankrashkin2008,Gitman2012,Juric2021}. Amongst those, there is a specially relevant $U(1)$ subfamily, with Hamiltonians $\hat{H} =4 E_{C_{s}} \left( \hat{n}_{\theta} - n_{g} \right)^{2} + V (\hat{\theta} )$. Here $\hat{n}_\theta$ is the momentum $-i\partial_\theta$ with periodic boundary conditions (PBC) and, therefore, integer spectrum\footnote{We signal operators by the use of circumflexes $\hat{}$.}. The parameters of the kinetic term are $E_{{C_s}}$, the overall energy scale, and $n_g$, which can be understood as the non-integer remainder of a magnetic flux through the ring, when the flux is expressed in $\Phi_0=h/2e$ magnetic flux quantum units (The particle being charged with charge $e$ and Planck's constant $h$).

At this point, we consider a concrete family of Hamiltonians of the form
\begin{equation}
\label{symHam}
\hat{H}_{2\theta} =4 E_{C_{s}} \left( \hat{n}_{\theta} - n_{g} \right)^{2}  - \lambda \cos{( 2 \hat{\theta} )}
\end{equation}
as the clearest example of a wider family with the same symmetry properties. Furthermore, there are proposals for its actual implementation with superconducting circuits. 

For all values of the parameters, this Hamiltonian has a $\mathbb{Z}_{2}$ symmetry generated by a rotation (translation on the circle) of angle $\pi$, $\hat{U}_{\pi} = e^{i \pi \hat{n}_{\theta}}$, such that in the position representation $\hat{U}_{\pi} |\theta \rangle = | \theta +\pi \rangle$ (always modulo $2\pi$). This is the remnant of the $SO(2)$ translation symmetry of the free ($\lambda=0$) case, explicitly broken by the potential.

The crucial point we want to make at this point is that there are two values of the $n_g$ parameter for which the group of symmetry is larger, namely at $n_{g} = 0$ and $n_{g} =1/2 $. To make this apparent, remember that, independently of the Hamiltonian, the ring has the geometric symmetry of reflection $\theta\to2\pi-\theta$. In the position representation this operation is realised by the unitary involution $\hat{U}_{P}|\theta \rangle = |2 \pi - \theta \rangle$. Momentum transforms by conjugation as $\hat{U}_P\hat{n}_\theta \hat{U}_P=-\hat{n}_\theta$, whence, on momentum eigenstates, $\hat{U}_{P} | n \rangle = |- n \rangle$. Clearly, this involution is a symmetry of $\hat{H}_{2\theta}$ for $n_g=0$, such that the symmetry group generated by $\hat{U}_\pi$ and $\hat{U}_P$ is the classical symmetry $\mathbb{Z}_2\times\mathbb{Z}_2$ of order four. The symmetry is unbroken in that the ground-state is non-degenerate. It is the remnant of the $O(2)$ group of the free $\lambda=0$ case.

At $n_{g} = 1/2$, there is a different involution that is a symmetry of the Hamiltonian, namely the (twisted) reflection symmetry $\hat{V}_{P}$, implemented in coordinate $\theta$ space as $\hat{V}_{P}|\theta \rangle =e^{-i\theta} |2 \pi - \theta \rangle$, and in momentum space as $\hat{V}_{P} | n \rangle = |1 - n \rangle$. Now, $\hat{V}_{P}^{2}=\hat{U}^{2}_{\pi} =\mathbb{I}$ and, crucially, $\hat{V}_{P} \hat{U}_{\pi} = -\hat{U}_{\pi} \hat{V}_{P}$. Thus the symmetry group is the dihedral $D_4$ group of order eight, a central extension of the classical symmetry. Notice that in the free case $\lambda=0$ the symmetry of the $n_g=1/2$ is also enhanced from the classical $O(2)$ group (See the Supplemental Material at \ref{sec:symm-free-part}, \ref{sec:symm-pres-potent} and \ref{sec:math-fram-proj} for further details \cite{SuppMat}). With interaction and the discrete $D_4$ symmetry group all energy levels, including the ground one, are necessarily degenerate: it is impossible for a state to be invariant under this unitary representation of the full $D_4$ group, because of the anti-commutation of $\hat{V}_P$ and $\hat{U}_\pi$. This degeneracy, at least twofold, is robust, in that it is independent of the precise form of the potential and its energy scales, as long as the potential is classically invariant under $\theta\to2\pi-\theta$ (expanded only in cosines) and under translation by $\pi$ (only even cosines).  It is important to note that there are characters (one-dimensional representations) of $D_4$, which would suggest the possibility of non-degenerate levels. However, the particular presentation at hand precludes this from happening.

Physically, one can argue that the degeneracy is caused by the destructive interference of tunnelling amplitudes from $0$ to $\pi$ in the positive direction and its reflected version from $2\pi$ to $\pi$. Those contributions coming from higher windings will also be organised according to chirality. As $n_g$ is a flux through the ring, it gives rise to an Aharonov--Bohm phase factor, and the relative phases of the direct and the reflected path are $e^{2i n_g\pi}$. Therefore the energy splitting due to tunnelling is, to all perturbative orders, proportional to $\cos(n_g\pi)$. Taking the possibility of winding into account, there will be a factor $\cos\left[(2N+1)n_g\pi\right]$ for the successive pairs of instanton contributions. The fluctuations around these instantonic solutions will have the same factor. Thus we observe that to all orders and winding numbers the interference pattern is fully destructive at $n_g=1/2$. See the Supplemental Material at \ref{sec:symmetry-tunneling} for further details \cite{SuppMat}.

\section{Superconducting circuit realisation}

\begin{figure}[th!]
\begin{center}
\includegraphics[width=0.5\textwidth]{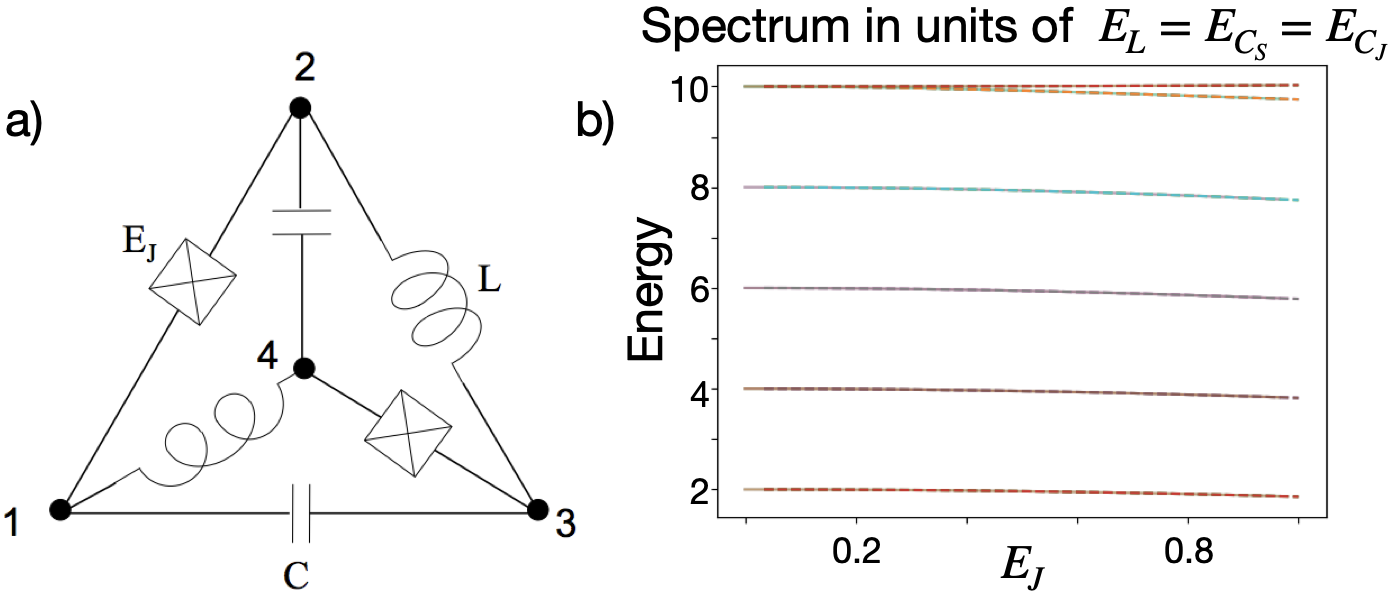}
\caption{a) Circuit diagram of the $0$-$\pi$ qubit. The circuit has one closed loop with four nodes connected by a pair of Josephson junctions $E_J$, a pair of capacitors $C$, and a pair of inductors $L$. b) Energy spectrum (\emph{every eigenvalue is two-fold degenerate}) for the complete Hamiltonian eq. \eqref{Hcomplete}, with energy scales $E_{C_{J}} = E_{C_{s}} = E_{L}$, parameters $n_{g}=1/2$, $\varphi_{\mathrm{ext}}=\pi$, and varying the energy scale $E_{J}$. At these values of $(n_{g} , \varphi_{\mathrm{ext}}) = (1/2 , \pi)$, the ``classical'' symmetry of the model is anomalous which appears in the degeneracy of all the levels in the spectrum of the Hamiltonian independently of the energy scales $E_{C_{J}}$, $E_{C_{s}}$, $E_{L}$, $E_{J}$.}
\label{fig: 0-pi circuit}
\end{center}
\end{figure}

In the following, after a review of the basic features of the $0$-$\pi$ qubit \cite{Kitaev2006,Brooks2013,Dempster2014,Groszkowski2018,Paolo2019,Gyenis2021}, we shall make apparent that it presents an anomalous symmetry, which leads to degeneracy of the fundamental level for all values of design parameters of the model. The corresponding operating regime is thus a candidate for a qubit protected with respect to design uncertainties.

The circuit of the $0$-$\pi$ qubit (see Fig.\ref{fig: 0-pi circuit} and also Supplemental Material at \ref{sec:explicit-break} \cite{SuppMat}) presents a pair of capacitors with capacitance $C$, a pair of inductors with inductances $L$ and a pair of Josephson junctions with Josephson energies $E_{J}$ and Josephson capacitances $C_{J}$. Moreover, the circuit is threaded by an external flux $\Phi_{\text{ext}}$. We introduce the dimensionless node flux variables $ \varphi_i=\frac{1}{\Phi_0}\int^t_{-\infty}V_i(\tau)d\tau$, where $V_i(\tau)$ is the electrostatic potential that depends on time for $i\in \{1,2,3,4\}$.

If we write the Hamiltonian in terms of these variables, the ``center-of-mass'' of the fluxes is seen to be a cyclic variable, and another linear combination represents an harmonic oscillator that (in the ideal case) decouples from the rest of the dynamics. Leaving aside those two variables, and using the combinations $\phi=(\varphi_2-\varphi_3)/2+(\varphi_4-\varphi_1)/2 $ and $\theta=(\varphi_2-\varphi_1)/2-(\varphi_4-\varphi_3)/2$ we are led to
\begin{equation}
\label{Hcomplete}
\begin{split}
H_{0\text{-}\pi}&=4E_{C_J} \hat{Q}^2_{\phi}+E_L\hat{\phi}^2+4E_{C_s} \left( \hat{n}_{\theta} - n_{g} \right)^2\\
&-2E_J\cos\hat{\theta}\cos \left(\hat{\phi}-\frac{\varphi_{\mathrm{ext}}}{2} \right)\,.
\end{split}
\end{equation}
Here $E_L=\frac{\Phi_0^2}{4\pi^{2}L}$ is the inductive energy, $E_{C_J}=\frac{e^2}{2C_J}$ ($E_{C_s}=\frac{e^2}{2C_s}$) denotes the charging energy conjugate to the $\phi$ $(\theta)$ mode with total capacitance $C_{J}$ $(C_{s})$, $\varphi_{\text{ext}}=\frac{2\pi \Phi_{\text{ext}}}{\Phi_0}$ is the external flux in natural variables, $n_{g}$ is the offset-charge bias due to the electrostatic environment, $ \hat{n}_{\theta}$ is the canonical charge operator (in units of $2e$) corresponding to the compact phase operator $\hat{\theta}$, with integer spectrum, and $ \hat{Q}_{\phi}$ is the canonical charge operator (in units of $2e$) corresponding to the non-compact phase operator $\hat{\phi}$.

The free ($E_J=0$) Hamiltonian is straightforwardly diagonalisable. The $\phi$ part is a harmonic oscillator $-4E_{C_J}\partial^2_{\phi}+E_L\phi^2 = 2 \sqrt{E_{C_J} E_L} \left( \hat{N}_{\phi} + 1/2 \right)$, with $\hat{\phi}= \left( \frac{4E_{C_J}}{E_L} \right)^{1/4} \frac{\hat{a}_{\phi}^{\dagger} + \hat{a}_{\phi} }{\sqrt{2}}$, and $\hat{N}_{\phi} = \hat{a}_{\phi}^{\dagger} \hat{a}_{\phi}$.

To better understand the spectrum also when $E_J\neq0$, we shall now study the symmetries of the Hamiltonian. Along the lines of the $\cos(2\theta)$ Hamiltonian above, there are reflection symmetries at $n_{g}=0$, namely $\hat{U}_{P}$, and at $n_{g}=1/2$, the twisted $\hat{V}_{P}$. There is no involutive translation symmetry involving only the $\theta$ variable. However, depending on the value of $\varphi_{\mathrm{ext}}$ there is one involving just $\phi$, or one with $\theta$ and $\phi$ together. Define $\hat{P}_{\phi} | \phi \rangle = | -\phi \rangle$. At $\varphi_{\mathrm{ext}}=0$ this is a symmetry of the Hamiltonian. Much more interestingly, at $\varphi_{\mathrm{ext}} = \pi$ the composition $\hat{U}_\pi\hat{P}_\phi$ is a symmetry of the Hamiltonian.  In fact, the kinetic terms, the harmonic potential, and the interaction term are each individually invariant under these symmetry operations, and the symmetry groups are the classical $\mathbb{Z}_2\times\mathbb{Z}_2$, generated by $\hat{U}_{P}$ and $\hat{U}_{\pi} \hat{P}_{\phi}$, at $(n_{g},\varphi_{\mathrm{ext}}) = (0,\pi)$, and the enhanced dihedral group $D_4$ generated by $\hat{V}_{P}$ and $\hat{U}_{\pi} \hat{P}_{\phi}$ at $(n_{g},\varphi_{\mathrm{ext}}) = (1/2,\pi)$.  The anomalous character of this last case is evident in the two-fold degeneracy of each energy level being kept for all values of the coupling energy scale $E_J$ (see Fig. \ref{fig: 0-pi circuit}). The other possibilities, at $(n_g,\varphi_{\mathrm{ext}})=(0,0)$ and $(n_g,\varphi_{\mathrm{ext}})=(1/2,0)$, give rise to $\mathbb{Z}_2\times\mathbb{Z}_2$, now generated by $\hat{P}_\phi$ and, correspondingly, $\hat{U}_P$ or $\hat{V}_P$.

As the symmetry, and hence the degeneracy of all energy levels, is maintained for all values of the energy parameters $\left\{E_{{C_J}},E_L,E_{{C_s}},E_J\right\}$ as long as the offset charge $n_g$ and external flux $\varphi_{\mathrm{ext}}$ are fixed at $(1/2,\pi)$, this system is a good candidate to provide us with useful qubits, robust with respect to design inaccuracies. In fact, when the oscillator frequency is much larger than the other frequency parameters of the model, the effective low-energy Hamiltonian is of the form investigated earlier and maintains the same symmetry and degeneracy. 

This statement is apparent using a perturbative Schrieffer--Wolff analysis \cite{SW1966,Altland2010,Bravyi2011} that preserves the full symmetry at each order (see Supplemental Material at \ref{sec:pert-schr-wolff} for details \cite{SuppMat}). Here we take the formal expansion parameters to be $E_{{C_s}}/\sqrt{E_{{C_J}}E_L}$ and $E_{{J}}/\sqrt{E_{{C_J}}E_L}$, both assumed to be of the same order with respect to the expansion. In other words, the low-energy sector will be understood as presenting no (dressed) $\phi$ excitation. The expansion can be computed for all values of $n_g$ and $\varphi_{\mathrm{ext}}$. At any given order, an effective potential for $\theta$ is generated, such that its Fourier series at $\varphi_{\mathrm{ext}}=\pi$ only includes even cosine terms, i.e., $\cos\left(2k\theta\right)$.

\begin{figure}[th!]
\begin{center}
\includegraphics[width=0.26\textwidth]{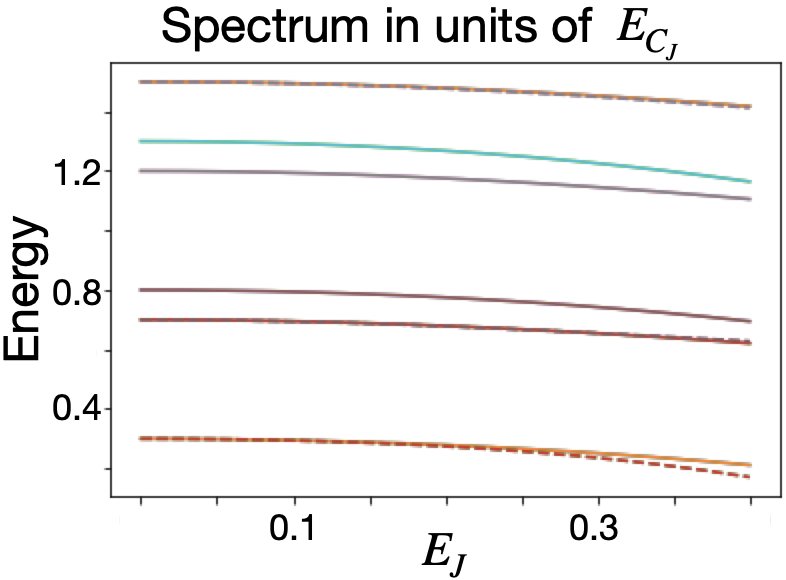}
\caption{ Comparison of the energy spectrum (\emph{every eigenvalue is two-fold degenerate}) for the complete Hamiltonian eq. \eqref{Hcomplete} (full lines) and the low-energy effective model eq. \eqref{Hpert} (dashed lines), in units of $E_{C_{J}}$ for $E_{C_{s}}=1/20$, $E_L=1/16$, $(n_{g},\varphi_{\mathrm{ext}}) = (1/2,\pi)$, and varying the energy scale $E_{J}$. The effective model assumes $N_{\phi} \sim 0$, which happens for the lines starting at the energies $\{0.3 , 0.7 , 1.5 \}$. The lines starting at the energies $\{ 0.8 , 1.2 \}$ have $N_{\phi} \sim 1$, while the one starting at $1.3$ has $N_{\phi} \sim 2$.}
\label{fig:pertur}
\end{center}
\end{figure}

At first order of the expansion and outside of the symmetry point $(n_{g},\varphi_{\mathrm{ext}}) = (1/2,\pi)$, the low-energy effective Hamiltonian reads
\begin{equation}
\label{eq:zerothapprBroken}
\begin{split}
& \hat{H}_{\mathrm{eff}}= \sqrt{E_{{C_J}}E_L}+4E_{C_s} \left( \hat{n}_{\theta} - n_{g} \right)^2 \\
&-2E_J\cos \hat{\theta} \cos \left( \frac{\varphi_{\mathrm{ext}}}{2} \right) e^{ - \frac{1}{2}\sqrt{ \frac{E_{C_J}}{E_L} }}.
\end{split}
\end{equation}
The effective potential term, to this order,  breaks the symmetry. In particular, it is not invariant under $\hat{U}_\pi$. However, it is exponentially suppressed when $E_{{C_J}} \gg E_{L}$. On further inspection of the expansion, one observes that all the symmetry-breaking terms will present this exponential attenuation. On the other hand, symmetry-preserving terms have an asymptotic (negative) power law behaviour in $\left(E_{{C_J}}/E_L\right)^{1/4}$. This limit has been studied in the literature, and its effect is to protect the low-energy sector from flux noise. Observe that the relevant parameter is an immittance for the harmonic oscillator part, and the limiting behaviour is determined by the dominance of  the kinetic (charge) term of the harmonic oscillator. Thus the system is highly delocalised in $\phi$, whence the impact that $\varphi_{\mathrm{ext}}$ can have is diminished. Furthermore, the dynamics is controlled by the effective $\theta$ Hamiltonian. In the democratisation of the flux parameter we have as a consequence that the enhanced symmetry $D_4$ is recovered outside of its critical value  by controlling just the charge offset.

We must stress that the effective Hamiltonian is valid outside of this kinetic dominance regime, if indeed the oscillator frequency is well separated from others.

As an example, the effective Hamiltonian at the point $(n_{g},\varphi_{\mathrm{ext}}) = (1/2,\pi)$ to third-order in the perturbative SW expansion, asymptotically when  $E_{C_{J}} \gg E_{L}$, is
\begin{equation}
\label{Hpert}
\begin{split}
\tilde{H}_{0\text{-}\pi}|_{N_{\phi}=0} &\xrightarrow[ E_{C_{J}} \gg E_{L} ] ~  \sqrt{E_{C_{J}} E_{L}} + 4 E_{C_{s}} \left( \hat{n}_{\theta} - \frac{1}{2} \right)^{2} \\
&-  \frac{E^{2}_{J}}{E_{C_{J}}} \left( 1 + \sqrt{\frac{E_{L}}{E_{C_{J}}}} \right) \cos^{2} \hat{\theta}  \\
&+ \frac{E^{2}_{J} E_{C_{s}}}{E^{2}_{C_{J}}} \left( 1 + 3 \sqrt{\frac{E_{L}}{E_{C_{J}}}} \right) \sin^{2}  \hat{\theta}\,,
\end{split}
\end{equation}
illustrated in Fig. \ref{fig:pertur}.

\begin{figure}[th!]
\begin{center}
\includegraphics[width=0.5\textwidth]{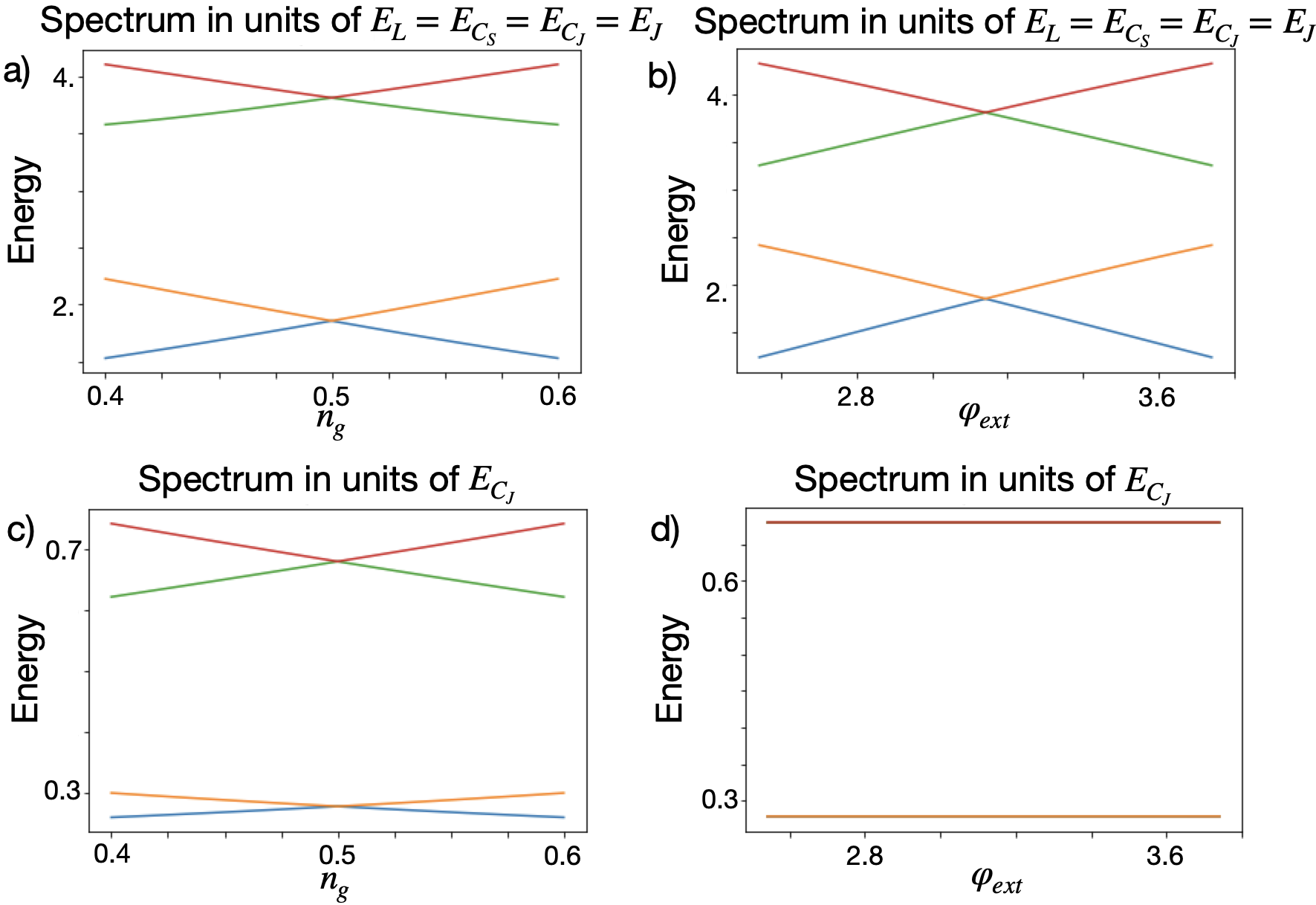}
\caption{The four lowest energy levels of the spectrum for the complete Hamiltonian eq. \eqref{Hcomplete} as a function of the offset charge $n_{g}$ (a) and c) panels) and external magnetic flux $\varphi_{\mathrm{ext}}$ (c) and d) panels). In the upper panels (a) and b)), the spectrum is in units of $E_{C_{J}} = E_{C_{s}} = E_{L}=E_{J}$; in the lower panels (c) and d)) the spectrum is in units of $E_{C_{J}}$ for $E_{C_{s}}=1/20$, $E_L=1/16$, and $E_{J}=1/5$. The lower panels illustrate  insensitivity to flux and charge noise for a kinetic dominance regime. }
\label{fig: parameter}
\end{center}
\end{figure}

Passing now to offset charge sensitivity (see also Supplemental Material at \ref{sec:charge-offs-sens} \cite{SuppMat}), observe that the effective Hamiltonian to third-order at $\varphi=\pi$ is a $\cos(2\theta)$ one, eqn. \eqref{symHam}. At all orders and values of the offset charge $n_g$ the discrete translation $\hat{U}_\pi$ will be a symmetry.  We can therefore separate the spectrum of the full effective Hamiltonian into an even and odd sector. These sectors will be connected by the $\hat{V}_P$ reflection, $|\mathrm{even}\rangle=\hat{V}_P|\mathrm{odd}\rangle$. Let us characterise the sensitivity by the dependence of the energy gap $E_1-E_0$ on $n_g$, as depicted in Fig. \eqref{fig: parameter}, around the $n_{g}=1/2$ point. Given the separation between even and odd states, and the Feynman--Hellman theorem, we desire to compute
\begin{equation}
\begin{split}
\frac{\partial\Delta}{\partial n_g}&= \frac{\partial E\left(\mathrm{gs},e\right)}{\partial n_g}-\frac{\partial E\left(\mathrm{gs},o\right)}{\partial n_g} \\
                                     &=8E_{{C_s}}\left(1-2\left\langle\mathrm{gs},e|\hat{n}_\theta|\mathrm{gs},e\right\rangle\right)\,.                         
\end{split}
\label{eq:gapderiv}
\end{equation}
Consider now the semiclassical limit $\lambda\gg E_{{C_s}}$,
\begin{equation*}
  \frac{\partial\Delta}{\partial n_g}=4E_{{C_s}} \pi \sqrt{ \frac{\lambda}{2 E_{C_{s}}} } e^{-\frac{\pi^{2}}{4} \sqrt{\frac{\lambda}{2E_{C_{s}}} }}\,.
\end{equation*}

Thus, a qubit built out of a $\cos(2\theta)$ would be protected with respect to charge noise at this symmetry point in this limit. From the point of view of the $0$-$\pi$ circuit, this suggests an operation regime for robust and protected qubits, given by the hierarchy
\begin{equation*}
  E_{{C_J}}^{2} \gg E_{{C_J}}  E_L \gg E_J^2 \gg E_{{C_J}} E_{{C_s}}\,.
\end{equation*}
This is determined by kinetic dominance, perturbative validity of the expansion, and semiclassicality of the effective Hamiltonian. In fact, the usefulness of the enhanced symmetry point extends far beyond this special regime, as Fig. \ref{fig: parameter} shows for the full $0$-$\pi$ Hamiltonian \eqref{Hcomplete}.

\section{Conclusion and outlook}
\label{sec:conclusions}

The main objective in this work was to understand the special symmetry properties of the superconducting circuit $0$-$\pi$ qubit. To achieve this goal, first we identify the full symmetry group of the model and show how it is realised at high-symmetry points. To have a closer characterisation of the low-energy dynamics, we perform a SW transformation respecting the symmetry of the model and study the degeneracy of the ground-state manifold and the action of the operators that split this degeneracy, explicit symmetry-breaking terms. It would be interesting to analyse how the anomalous realisation of the symmetry is modified in more realistic models where the noise and perturbations are described by open quantum systems. 

\section*{Acknowledgments}

We thank A. Parra-Rodriguez for valuable insights on the circuit realisation during the early stages of the manuscript. E.R. thanks the QuantERA projects QTFLAG and T-NiSQ. E.R. and I.L.E. acknowledge support of the Basque Government grant IT986-16.

\pagebreak
\newpage

\bibliographystyle{JHEP}
\providecommand{\href}[2]{#2}
\begingroup
\raggedright

\endgroup

\pagebreak
\newpage

\widetext
\begin{center}
\textbf{\large Role of anomalous symmetry in $0$-$\pi$ qubits}

I.L. Egusquiza, A. I\~niguez, E. Rico, A. Villarino
\end{center}

\section{Supplemental Material}

\subsection{Symmetries for the free particle}
\label{sec:symm-free-part}

Let us consider a free classical particle moving on the ring $S^1$, parameterised with the interval $[0,2\pi)$. The dynamical variable will be denoted as $\theta$. To better connect with the physics of the superconducting circuits we want to describe, we choose to write the Lagrangian as
\begin{equation}
\label{eq:lagclassical}
L=\frac{1}{16\epsilon}\dot{\theta}^2+ n_g\dot{\theta}\,.
\end{equation}

The classical Hamiltonian derived from this Lagrangian is determined to be
\begin{equation}
\label{eq:hamclassical}
H=4\epsilon\left(p-n_g\right)^2\,.
\end{equation}
Please observe that the second term of the Lagrangian, being a total derivative, has no effect on the equations of motion, and could readily be eliminated by moving to a different reference frame. It can, however, be used to determine a particular quantisation in the path integral formalism. In this text, we shall make specific choices of quantisation, as we shall make clear later. The path integral approach will be relevant for a physical argument regarding destructive interference of tunnelling amplitudes.

We desire to describe motion on the ring, and thus it is convenient to relate those symmetries of the equations of motion, common to free particles in other configuration spaces, to the geometric symmetries of the case at hand, by using $\theta\mod 2\pi$ to identify points on the ring. 

There are clearly identifiable symmetries in this classical system, with corresponding conserved quantities. The first obvious one is rotation symmetry, since $\theta$ only appears through its derivative, and the associated conserved quantity is the canonical momentum $p$, as expected. Secondly, we have reflection symmetry, that we choose to present as symmetry by inversion around the $2\pi$ point, $ \theta\to2\pi-\theta$. These operations complete the full invariance group of the circle, $O(2)$.

Coming now to the quantised description of the system, it is imperative to bear in mind that first, quantisation is not necessarily unique, and, second, that not all classical symmetries survive the process of quantisation in exactly the same form. This last fact is reflected in the concept of anomaly, that has proven so extremely fruitful in high energy and condensed matter physics. Both of these points are relevant to the case of a particle on the circle.

Regarding the first point, there exists a $U(2)$ family of inequivalent quantisations of the Hamiltonian \eqref{eq:hamclassical}. Inside this $U(2)$ family there exists a $U(1)$ subfamily in which the quantised Hamiltonian operator can be understood in the form of \eqref{eq:hamclassical}, with $p$ a self-adjoint quantum canonical momentum. These momenta are inequivalent, since their spectra are different, and similarly the quantisations of the Hamiltonian are not isospectral. In fact, some of those quantised Hamiltonians present edge states with negative energies. The analysis we will carry out here could be presented in direct relation to these facts; we choose however to select one concrete canonical quantisation and stick to it throughout, so as not to confuse the issues \cite{Bonneau2001,Pankrashkin2008,Gitman2012,Juric2021}. 

Thus, we will henceforward consider \emph{only}  the quantum canonical momentum determined by $-i\partial_\theta$ with periodic boundary condition (PBC) $\psi(0)=\psi(2\pi)$ . Since its spectrum is the set of integers, we shall denote it by $\hat{n}$. Regarding notation, we will only insist on making the explicit distinction between operator and c-number for those cases in which we shall use the same character for them, as in $\hat{n}$ and integer $n$ or the position operator on the circle, $\hat{\theta}$, and a position coordinate $\theta$, while the quantum Hamiltonians will not carry a special operator mark.

Our first task, therefore, is to study the symmetries of the quantum Hamiltonians
\begin{equation}
\label{eq:qham}
H=4\epsilon\left(\hat{n}-n_g\right)^2\,.
\end{equation}
Have in mind that the change $n_g$ to $1+n_g$  does not change the spectrum (and  neither does the shift of $n_g$ by any integer). In what follows, the \emph{offset number} $n_g$ will be restricted to a value in $[0,1)$.

The eigenstates of this free Hamiltonian are those of the PBC canonical momentum $\hat{n}$, denoted by $|n\rangle$ in what follows, that in the position representation have wave-functions
\begin{equation}
\label{eq:psin}
\psi_n(\theta)=\langle\theta|n\rangle=\frac{1}{\sqrt{2\pi}}e^{i n\theta}\,.
\end{equation}
Unless explicitly indicated, $n \in \mathbb{Z}$ will take values in the set of integers. Notice that, in keeping with $|n\rangle$ being an eigenstate of momentum, the probability density $|\psi_n(\theta)|^2=1/2\pi$ is homogeneous on the circle for all $n$.

Let us now examine the symmetries of this family of Hamiltonians. By construction there will be rotation symmetry. Namely, rotations will be generated by canonical momenta, and choosing $\hat{n}$ as that generator, it commutes with the Hamiltonian. The abelian group of rotations is in this case $U(1)$, isomorphic to $SO(2)$, with elements
\begin{equation}
\label{eq:u1n}
\hat{U}_R(\alpha)= e^{-i \alpha\hat{n}}\,,
\end{equation}
where the parameter $\alpha$ belongs to the interval $[0,2\pi)$.

Discrete symmetries are rather different. Reflection around an axis is a transformation of configuration space, and is therefore implemented in the position basis of Hilbert space by
\begin{equation}
\label{eq:parityHilbert}
\hat{U}_P|\theta\rangle=|2\pi-\theta\rangle\,.
\end{equation}
It follows that
\begin{equation}
\label{eq:parityonnumber}
\hat{U}_P|n\rangle=|-n\rangle\,,
\end{equation}
making $\hat{U}_P$ an involution, $\hat{U}_P^2=1$, and it follows that
\begin{subequations}
\begin{align}
\hat{U}_P\hat{n}\hat{U}_P&=\hat{U}_P\left(\sum_{n\in\mathbb{Z}}n|n\rangle\langle n|\right)\hat{U}_P=-\hat{n}\,,\label{eq:upn}\\
\hat{U}_P H \hat{U}_P&=4\epsilon\left(\hat{n}+n_g\right)^2\,.\label{eq:uph}
\end{align}
\end{subequations}
This is the crucial point that needs careful analysis, so let us make a distinction between the case $n_g=0$ and other situations, $n_g\in(0,1)$.

\subsubsection{$n_g=0$: the $O(2)$ group }
\label{sec:n_g=0}

In the case $n_g=0$ it is immediate to observe that the Hamiltonian commutes with $\hat{U}_P$. Furthermore, since
\begin{equation}
\label{eq:utup}
\hat{U}_P\hat{U}_R(\alpha)\hat{U}_P=\hat{U}_R^\dag(\alpha)=\hat{U}_R(2\pi-\alpha)\,,
\end{equation}
we have a faithful unitary representation of $O(2)$ that commutes with the Hamiltonian $H=4\epsilon\hat{n}^2$, and the quantum system presents the same symmetries as the original classical one.

\subsubsection{$n_g\neq0$, the fixed point $n_g=1/2$ and the $\mathrm{Pin}(2)$ group.}
\label{sec:n_gneq0-fixed-point}

Now, since $\hat{U}_P H \hat{U}_P=4\epsilon\left(\hat{n}+n_g\right)^2\neq H$, $H$ does not commute with $\hat{U}_P$ for $n_g\neq0$. Even so, in this
case
\begin{align}
\hat{U}_P H \hat{U}_P&= 4\epsilon\left(\hat{n}+n_g\right)^2\\
&= 4\epsilon\left[\hat{n}+1-\left(1-n_g\right)\right]^2\\
&= 4\epsilon e^{-i\hat{\theta}}\left[\hat{n}-\left(1-n_g\right)\right]e^{i\hat{\theta}}\,.\label{eq:unitequiv}
\end{align}
Here we have used the fact that $\hat{n}$, with integer spectrum, is conjugate to the position operator on the circle, $\hat{\theta}$. This means that on the $|n\rangle$ basis
\begin{equation}
\label{eq:eit}
e^{i k\hat{\theta}}=\sum_{n\in\mathbb{Z}}|n+k\rangle\langle n|
\end{equation}
for integer $k \in \mathbb{Z}$.

Eq. \eqref{eq:unitequiv} tells us that the free Hamiltonian with offset number $n_g\in(0,1)$ (open set) is unitarily equivalent (and therefore isospectral) to the free Hamiltonian with offset number $1-n_g\in(0,1)$.  We must stress that $(\hat{n}-n_g)^2$ and $(\hat{n}-n'_g)^2$ have different spectra for generic $n_g\neq n_g'$, and describe therefore different dynamics. It follows that for generic $n_g$ the reflection transformation implemented by the unitary $\hat{U}_P$ is \emph{not} a symmetry of the Hamiltonian: the classical $O(2)$ symmetry is broken down to $SO(2)$ rotation symmetry.

There is however a special point, $n_g=1/2$, which corresponds to the fixed point of the transformation $n_g\to1-n_g$. In this case the Hamiltonian is \emph{not} invariant under the action of $\hat{U}_P$, but is transformed by it into a unitarily equivalent operator.  Hence, the unitary transformation
\begin{equation}
\label{eq:vp}
\hat{V}_P= e^{i\hat{\theta}}\hat{U}_P
\end{equation}
\emph{is} a symmetry of the Hamiltonian with $n_g=1/2$. Notice that this unitary is actually hermitian, since
\begin{equation}
\label{eq:upeialphatheta}
\hat{U}_P e^{i\alpha \hat{\theta}}\hat{U}_P=e^{2\pi i \alpha} e^{-i\alpha\hat{\theta}}\,,
\end{equation}
whence
\begin{equation}
\label{eq:upitheta}
\hat{U}_P e^{i\hat{\theta}} \hat{U}_P= e^{-i\hat{\theta}}\,,
\end{equation}
from which
\begin{equation}
\label{eq:vphermitian}
\hat{V}_P^\dag=\hat{U}_Pe^{-i\hat{\theta}}= \hat{U}_P^2 e^{i\hat{\theta}}\hat{U}_P= e^{i\hat{\theta}}\hat{U}_P=\hat{V}_P\,.
\end{equation}
We see that $\hat{V}_P$ is therefore also an involution, $\hat{V}_P^2$. Acting on the canonical PBC momentum basis, we have
\begin{equation}
\label{eq:vponn}
\hat{V}_P|n\rangle=|1-n\rangle\,,
\end{equation}
from which the unitarity and involutive aspect of the operator are clearly identified.

In summary to this point, for $n_g=1/2$, the Hamiltonian has a continuous $U(1)$ symmetry, with group elements $\hat{U}_R(\alpha)=e^{-i\alpha\hat{n}}$, and an involutive symmetry $\hat{V}_P$. Crucially, and in contradistinction to eq. \eqref{eq:utup}, now we have
\begin{equation}
\label{eq:vput}
\hat{V}_P \hat{U}_R(\alpha) \hat{V}_P = e^{-i\alpha} \hat{U}_R^\dag(\alpha)\,.
\end{equation}
This can be written in an alternative form, that can guide us to a better understanding of the underlying structure,
\begin{equation}
\label{eq:vphalfn}
\hat{V}_P e^{-i \alpha\left(\hat{n}-1/2\right)}=e^{i \alpha\left(\hat{n}-1/2\right)}\hat{V}_P\,.
\end{equation}
This expression is analogous to \eqref{eq:utup}, with one crucial difference. We now have a family of unitaries with generator $\hat{n}-1/2$, instead of $\hat{n}$. This means that we actually have a double cover of $U(1)$, since we need to go from $\alpha=0$ to $\alpha=4\pi$, instead of $2\pi$, to recover the unit. Clearly,
\begin{equation}
\label{eq:2pivalue}
e^{-i 2\pi\left(\hat{n}-1/2\right)}=-1\,.
\end{equation}
Thus the symmetry is a double cover of $O(2)$. This double cover is termed $\mathrm{Pin}(2)$.

For clarity, let us discuss this $\mathrm{Pin}$ group in abstract terms, and then show that there is indeed a faithful representation of the Pin group in unitaries on the Hilbert space that commutes with the Hamiltonian for the case $n_g=1/2$, thus showing that this is the \emph{enhanced} symmetry group in this situation.

An explicit form for $\mathrm{Pin}(2)$ is given by $F(0)=1$ and the composition laws
\begin{eqnarray}
\label{eq:pin2table}
F(x) F(y)&=& F(x+y)\,,\nonumber\\
E(x) E(y)&=& F(x-y+\pi)\,,\nonumber\\
E(x) F(y)&=& E(x-y)\,,\\
F(x) E(y)&=& E(x+y)\,.\nonumber
\end{eqnarray}
Here $x$ and $y$ are in the interval $[0,2\pi)$ and addition and subtraction are to be understood modulo $2\pi$. That is, $F(2\pi)=F(0)=1$. From the first line of this table, one reads the embedding of an $SO(2)$ group inside this one. Next, let us point out some quirks of this group:
\begin{eqnarray}
\label{eq:eprops}
E(x)^2&=& F(\pi)\,,\nonumber\\
E(x)^4&=& F(2\pi)=1\,,\nonumber\\
E(x)^{-1}&=& E(x+\pi)=E(x) F(\pi)=F(\pi) E(x)\,,\\
F(x)^{-1}&=& F(-x)\,.\nonumber
\end{eqnarray}

We now have the explicit representation of this group
\begin{align}
\label{eq:reppin2}
F(x)&= e^{-2i x\left(\hat{n}-1/2\right)}\,,\\
E(0)&= i \hat{V}_P\,,\\
E(x)&= F(x)E(0)= i e^{-2i x\left(\hat{n}-1/2\right)}\hat{V}_P\,.
\end{align}
In order to check that the multiplication table of eq. \eqref{eq:pin2table} is represented one must make use of eq. \eqref{eq:vphalfn}. It is also easy to check that the representation is faithful.

We have therefore shown that the symmetry group of the Hamiltonian for the free particle on the circle with PBC canonical quantisation  can be
\begin{itemize}
\item identical to the classical symmetry $O(2)$ when the offset number $n_g=0$;
\item enhanced to the larger $\mathrm{Pin}(2)$ group when the offset number is $n_g=1/2$; 
\item broken down to $SO(2)$ to other values of the offset number $n_{g} \in (0,1)$.
\end{itemize}

\subsection{Symmetries in presence of a potential}
\label{sec:symm-pres-potent}

We are considering motion on a circle; we therefore require potentials that can be expressed with periodic functions of period $2\pi$. A general periodic function will break fully the symmetries of the circle, both reflections and rotations. Therefore, let us consider only periodic functions that are invariant under the reflection $\theta\to2\pi-\theta$, i.e. those that admit a cosine Fourier series,
\begin{equation}
\label{eq:vtheta}
V(\theta)=\sum_{k=1}^\infty V_k \cos(k\theta)\,.
\end{equation}
The classical rotation symmetry of the free Lagrangian is broken by the presence of these terms. However, there can be residual discrete rotation symmetries for some potentials. For example, assume that $V_{2k+1}=0$ (only even terms remain).  In such a case, a rotation by $\pi$ (modulo $2\pi$), $\hat{U}_\pi=e^{i\pi\hat{n}}$, leaves the potential unchanged. This discrete rotation is an involution. In other words, the identity and this translation form a $\mathbb{Z}_2$ group, and the rotation group $SO(2)$ is broken down to this discrete group.

Therefore, in the case $n_g=0$, the full symmetry group is broken, and the breaking pattern is $O(2)\to G_0=\mathbb{Z}_2\times\mathbb{Z}_2$, with
\begin{equation}
\label{eq:gzero}
G_0=\left\{1,\hat{U}_P,\hat{U}_\pi,\hat{U}_P \hat{U}_\pi\right\}
\end{equation}
as a set. On the eigenbasis, the generators act as
\begin{align}
\label{eq:zed2timeszed2}
\hat{U}_P|n\rangle &= |-n\rangle\,,\\
\hat{U}_\pi|n\rangle&= (-1)^n|n\rangle\,.
\end{align}

Much more interesting is the case $n_g=1/2$, in which the symmetry-breaking in the presence of a potential with only even $k$ terms is $\mathrm{Pin}(2)\to G_\pi=D_4$, the symmetry group of the square. This group presents two generators  such that $D_4=\left\langle a,b : a^4=b^2=1,\, ab = b a^{-1}\right\rangle$. Notice that, using this notation, $a^2$ and $b$ are involutions that generate a $\mathbb{Z}_2\times \mathbb{Z}_2$ subgroup.

Let us analyse the realisation of this group in detail. We have seen earlier that the role of the reflection $\hat{U}_P$ is assumed by the twisted reflection $\hat{V}_P$ to keep invariant the kinetic term $4\epsilon\left(\hat{n}-n_g\right)^2$. This twisted reflection anti-commutes  with rotation by an angle  $\pi$, eq. \eqref{eq:vput}. On the eigenbasis, they act as 
\begin{align}
\label{eq:d4gens}
\hat{V}_P|n\rangle&= |1-n\rangle\,,\\
\hat{U}_\pi|n\rangle &= (-1)^{n}  |n\rangle\,.
\end{align}
Furthermore, since $\hat{V}_P=e^{i\hat{\theta}}\hat{U}_P$, the twisted reflection also commutes with the potential term. Next, observe that the identification 
\begin{align}
\label{eq:d4assignment}
a&\to \hat{U}(a)=e^{i\pi\hat{n}}e^{i\hat{\theta}}\hat{U}_P\,,\\
b&\to \hat{U}(b)=\hat{V}_P=e^{i\hat{\theta}}\hat{U}_P
\end{align}
gives us a presentation of the dihedral group at hand.

Explicitly, acting on the PBC momentum eigenbasis
\begin{align}
\label{eq:d4action}
\hat{U}(a)|n\rangle&= (-1)^{n+1}|1-n\rangle\,,\\
\hat{U}(b)|n\rangle&= |1-n\rangle\,.
\end{align}
A faithful representation of $D_4$ is given by $a\to i\sigma^y$ and $b=\sigma^x$, and we see that on $\{|n\rangle,|1-n\rangle\}$ doublets $\hat{U}(a)$ and $\hat{U}(b)$ act as $(-1)^n i\sigma^y$ and $\sigma^x$ respectively.

In summary, we have proven that the Hamiltonians of the form
\begin{equation}
\label{eq:co2thetahamscanonical}
H= 4\epsilon \left(\hat{n}-\frac{1}{2}\right)^2 -\sum_{k=1}^\infty V_{2k}\cos\left(2k\hat{\theta}\right)
\end{equation}
have a symmetry $D_4$ which is an \emph{enhancement} with respect to the case $n_g=0$, which presents the same discrete group of symmetries as the classical Lagrangian, namely $\mathbb{Z}_2\times\mathbb{Z}_2$.

There is one feature of this structure that must be stressed, namely the twofold degeneracy not just of excited levels, but of the ground-level as well. For clarity, the dihedral group $D_4$ of order eight does have one dimensional representations while the minimal faithful representation is two-dimensional. From the symmetry group perspective there would be no obstacle to the ground-state being non-degenerate. In fact, under some fairly general assumptions, the ground-state of a quantum particle must be non-degenerate, and therefore must be invariant under the symmetry group of the Hamiltonian. Indeed, that is the general expectation for quantum particles. Those assumptions are not met for the Hamiltonians under study, and, as stated, the ground-state is degenerate.

Let us now provide an algebraic proof of the degeneracy of the ground state: the unitary representation that realises the symmetry group gives us the anticommutation
\begin{equation}
  \label{eq:anticommute}
  \hat{U}_\pi \hat{V}_P=-\hat{V}_P \hat{U}_\pi\,,
\end{equation}
which cannot be realised in any one-dimensional representation. In other words, the actual representation of the group also must be understood in the context of the algebra of operators, with nontrivial $\hat{U}_\pi $ and $\hat{V}_P$, whence the conclusion. 

\subsection{Symmetry and tunneling}
\label{sec:symmetry-tunneling}

Let us consider a physical, semiclassical argument for the degeneracy of all energy levels at the special $n_g=1/2$ point. We shall make the argument for the potential $-V_2\cos(2\theta)$, and it can readily be generalised to the whole class of Eq. \eqref{eq:co2thetahamscanonical}.

The minima of the potential are located at $\theta=0$ and $\theta=\pi$. In a semiclassical analysis, any splitting between the lowest energy level and the next one, breaking degeneracy, will come about because of tunnelling. The first approximation to tunnelling amplitudes will be given by instantons from one minimum to the other. Notice that there are \emph{two} instantonic paths from 0 to $\pi$: the path along positive angles from 0 to $\pi$, and the path along negative angles from $0$ to $-\pi$. Alternatively, the second path is the image under $\hat{U}_P$ reflection of the first one, and goes from $2\pi$ to $\pi$.

The amplitudes are computed in first approximation via the euclidean action. The crucial mathematical point to this argument is that because of the $n_g\dot{\theta}$ term the euclidean action is actually complex,
\begin{equation}
  \label{eq:euclideanaction}
  S_E[\theta(t)]=\int_{t_{\mathrm{i}}}^{t_{\mathrm{f}}}\mathrm{d}t\,\left[\frac{1}{16\epsilon}\dot{\theta}^2-i n_g\dot{\theta}+V(\theta)\right]\,.
\end{equation}
The euclidean equations of motion are thus
\begin{equation}
  \label{eq:euclideaneom}
  \ddot{\theta}-8 \epsilon V'(\theta)=0\,.
\end{equation}
Notice that under the reflection $\theta\to2\pi-\theta$ this equation of motion is invariant. Let us denote with $\theta_+(t)$ the instanton solution from $0$ to $\pi$ and with $\theta_-(t)$ the reflected path. We see that
\begin{align}
  \label{eq:instantaction}
 \mathrm{Re}\left\{S_E\left[\theta_+\right]\right\}&=
                                                     \mathrm{Re}\left\{S_E\left[\theta_-\right]\right\}\,,\\
 \mathrm{Im}\left\{S_E\left[\theta_+\right]\right\}&=- 
 \mathrm{Im}\left\{S_E\left[\theta_-\right]\right\}= -n_g\pi\,.
\end{align}
Therefore, the total tunnelling amplitude in this approximation is
\begin{align}
  \label{eq:amplitude}
  K_E(0,\pi)&\approx e^{-S_E\left[\theta_+\right]}+e^{-S_E\left[\theta_-\right]}\\
  &= e^{-\mathrm{Re}\left\{S_E\left[\theta_+\right]\right\}}\cos\left(n_g\pi\right)\,.
\end{align}
We see therefore that at the special symmetry point the cosine term is zero. Furthermore, even though this computation is approximate, the result goes beyond this order, since the relative phase factors will always sum up to the cosine term.

More physically, since we can understand the charge-offset parameter as a magnetic flux threading the ring, there is a relative $e^{2i n_g\pi}$ factor, and there is full \emph{destructive interference}  when $n_g=1/2$.

The energy splitting associated with the tunnelling is then proportional to $\cos(n_g\pi)$, and we recover degeneracy at the special symmetry point $n_g=1/2$, for all values of the potential strength. We also see that the splitting is linear in $n_g-1/2$ close to the symmetry point.

In the analysis to this point we have only considered the direct and reflected path, without the possibility of winding. For compactness, define
\begin{align}
  \label{eq:RIdef}
  R& = \mathrm{Re}\left\{S_E\left[\theta_+\right]\right\}\qquad
     \mathrm{and}\nonumber\\
  I&= \mathrm{Im}\left\{S_E\left[\theta_+\right]\right\}= -n_g\pi\,.  
\end{align}
Consider the instanton from $0$ to $3\pi$, i.e., with winding number $N=+1$. The euclidean action corresponding to it is $3(R+iI)$. Associating the instantonic  solution $\theta_-$ to winding number $N=0$, and the path from $0$ to $-3\pi$ with $N=-1$ and so on, we compute the corresponding actions as being
\begin{equation}
  \label{eq:actN}
  S_N=\left(2|N|+1\right)\left[R+i\,\mathrm{sign}(N) I\right]\,.
\end{equation}
The corresponding amplitudes sum as
\begin{align}
  \label{eq:ampsum1}
K_E(0,\pi)&\approx e^{-S_E\left[\theta_+\right]}+
            e^{-S_E\left[\theta_-\right]}+\sum_{N=1}^\infty
            \left(e^{-S_N}+e^{-S_{-N}}\right)\nonumber\\
  &= \sum_{N=0}^\infty e^{-(2N+1)R}\left[e^{-(2N+1) i I}+ e^{(2N+1)i
    I}\right]\nonumber\\
  &=  \frac{2\sinh R\,\cos I}{\cosh(2R)-\cos(2I)}\,.
\end{align}
Again, we see destructive interference at $n_g=\pi$, and that the splitting is linear in $n_g-1/2$ close to the symmetry point.

In fact, this result can be made more precise by arguing in terms of the full path integral. The point is that for every path $\theta(t)$ from $0$ to $\pi$ there is a reflected path $P\theta(t)$, given by $P\theta(t)=-\theta(t)$. Thus the real and imaginary  parts of the  euclidean action for $\theta(t)$ and its reflection are related by
\begin{align}
  \label{eq:generalpathaction}
  \mathrm{Re}\left\{S_E\left[P\theta(t)\right]\right\}&=
                                                       \mathrm{Re}\left\{S_E\left[\theta(t)\right]\right\}\,,\\
  \mathrm{Im}\left\{S_E\left[P\theta(t)\right]\right\}&=
  -\mathrm{Im}\left\{S_E\left[\theta(t)\right]\right\}\,.
\end{align}
Furthermore, the imaginary part of the euclidean action for any path is, from Eq. \eqref{eq:euclideanaction}, the total angle traversed multiplied by $-n_g$. For all paths from $0$ to $\pi$ this will be $(2n+1)n_g\pi$, with $n$ an integer (we refrain from naming it as the winding number, since we have used a slightly different use above - for positive $n$ and $N$, $n$ and $N$ coincide, while for negative $n$ they are related by $n=-1+N$, since the path $\theta_-$ has $n=-1$). Classifying paths according to this integer, we compute
\begin{align}
  \label{eq:pathintcomp}
  K_E(0,\pi)&=\int_{\theta_i=0}^{\theta_f=\pi}\mathcal{D}\theta\,
              e^{-S_E\left[\theta\right]}\nonumber\\
  &=\frac{1}{2}\left[\int_{\theta_i=0}^{\theta_f=\pi}\mathcal{D}\theta\,
              e^{-S_E\left[\theta\right]}+ \int_{\theta_i=0}^{\theta_f=\pi}\mathcal{D}P\theta\,
    e^{-S_E\left[P\theta\right]}\right]\nonumber\\
    &=\frac{1}{2}\int_{\theta_i=0}^{\theta_f=\pi}\mathcal{D}\theta\,\left[
      e^{-S_E\left[\theta\right]}+e^{-S_E\left[P\theta\right]}\right]\nonumber\\
  &=\int_{\theta_i=0}^{\theta_f=\pi}\mathcal{D}\theta\,
    e^{-\mathrm{Re}\left\{S_E\left[\theta\right]\right\}}\cos\left[\mathrm{Im}\left(S_E\left[\theta\right]\right)\right]\nonumber\\
  &=\sum_{n\in\mathbb{Z}}\int_{\theta_i=0}^{\theta_f=\pi}\mathcal{D}\theta_n\,
              e^{-\mathrm{Re}\left\{S_E\left[\theta\right]\right\}}\cos\left[(2n+1)n_g\pi\right]\,.
\end{align}
We see again the same result: the tunnelling amplitude from one minimum to the other vanishes at the special symmetry point $n_g=1/2$, and thus the lowest energy splitting as well. Notice furthermore that this result applies more generally and not just to the $\cos(2\theta)$ potential: it holds if indeed $0$ and $\pi$ are the potential minima and the reflection symmetry is present.

In the case of $V(\theta)=-\lambda\cos(2\theta)$ we can give some explicit expressions for the classical contribution. For definiteness, it is necessary to subtract from the euclidean Lagrangian the value of the potential at 0, i.e., $V(\theta)=\lambda-\lambda\cos(2\theta)$. The following results are known in the context of the sine--Gordon equation; notice that in our case the variable is compact. The euclidean equation of motion \eqref{eq:euclideaneom} becomes
\begin{equation}
  \label{eq:eomcos}
  \ddot\theta - 16\epsilon\lambda\sin(2\theta)=0\,.
\end{equation}
The instanton solution from $0$ to $\pi$ is, setting aside moduli
\begin{equation}
  \label{eq:thetainst}
  \theta_I(t)=2\mathrm{arctan}\left[e^{4t\sqrt{2\epsilon\lambda}}\right]\,.
\end{equation}
The real part of the euclidean action corresponding to this instanton is
\begin{equation}
  \label{eq:realinstanton}
  \mathrm{Re}\left\{S_E\left[\theta_I\right]\right\}=\sqrt{\frac{2\lambda}{\epsilon}}\,.
\end{equation}
This justifies our assertion that in the transmon limit \cite{{Koch2007}}, $\lambda\gg \epsilon$, tunnelling is exponentially suppressed. Notice that this assertion is independent of operation at a sweet spot $n_g=1/2$ or not. Naturally enough, here we have only considered the classical action, without an analysis of fluctuations, but the exponential suppression is a global factor of the asymptotic series.

\subsection{Mathematical framework: Projective representations and central extensions}
\label{sec:math-fram-proj}
In this section \cite{Moore2021,Arovas2021,Hall2013}, we would like to understand why the dihedral group $D_4$ appears as the complete symmetry group at $n_g=\frac{1}{2}$, instead of the abelian group $\mathbb{Z}_2 \times \mathbb{Z}_2$. In the classical form, we identify two symmetries $a$ and $b$, the rotation by $\pi$ and the reflection along the $OX$-axis. These are both involutions and generate an abelian group $G=\mathbb{Z}_2 \times \mathbb{Z}_2$. However in the quantum setting, the symmetric group at hand is no longer abelian. \\

Let's consider the abelian group $G=\mathbb{Z}_2 \times \mathbb{Z}_2=\langle 1,a,b,ab\rangle$ and we consider the following representation of $G$ by associating $a\to \sigma^z, b\to\sigma^x$ and $ab\to \sigma^z\sigma^x$. This is not an ordinary representation of $G$ as it is not a homomorphism; that is, for all $g_i,g_j\in G$ the following property is not satisfied:
\[
U(g_i)U(g_j)=U(g_ig_j).
\]

However, if we relax its definition and ask for the multiplication rule to be preserved up to a phase $e^{i\gamma(g_i,g_j)}$, then we have a projective representation $U$:
\[
U(g_i)U(g_j)=e^{i\gamma(g_i,g_j)}U(g_ig_j).
\]
And this is exactly what we have above since although $e^{i\gamma(g_i,g_j)}=1$ for $\gamma(g_i,g_j)=\gamma(a,b)=\gamma(a,ab)=\gamma(ab,b)$, $e^{i\gamma(g_i,g_j)}=-1$ for 
\[
\gamma(g_i,g_j)=\gamma(b,a)=\gamma(ab,a)=\gamma(b,ab).
\]

The associativity property is still required for a projective representation and so $\gamma(g_i,g_j)$ must satisfy the consistency condition $\gamma(g_i,g_j g_k) + \gamma(g_j,g_k) = \gamma(g_ig_j,g_k) + \gamma(g_i,g_j)$. This phase $e^{i\gamma(g_i,g_j)}$ is called the \textit{Schur multiplier} of the projective representation. Two equivalent projective representations have multipliers differing in a trivial multiplier and the equivalence class of multipliers is what we call today the second cohomology group of $G$, $H^2(G,\mathbb{C}^*)$. In the case above, $H^2(\mathbb{Z}_2 \times \mathbb{Z}_2,\mathbb{C}^*)=\mathbb{Z}_2$. \\

Abelian groups do have corresponding one-dimensional ordinary representations but we have just seen that they do actually accept two-dimensional projective representations. We will be able to lift this projective representation to an ordinary one of a larger group $\tilde{G}$. $\tilde{G}$ is then called a central extension of $G$. Specifically, a central extension $\tilde{G}$ of a group $G$ by an abelian group $A$ is an exact sequence of homomorphisms; i.e., 
\[
1 \xrightarrow ~ A \xrightarrow{\iota} \tilde{G} \xrightarrow{\mathcal{\pi}} G \xrightarrow ~ 1
\]
such that the kernel of every map in the sequence is the image of the map which precedes it and that $\iota$ is injective and $\pi$ is surjective. The first map in the sequence is just the injection of the one-element group $\{1\}$ to the identity in $A$ and the last map is the trivial surjection onto $\{1\}$. Besides, $\iota(A)\subseteq Z(\tilde{G})$, as it is a central extension. We then say that $G$ is lifted to $\tilde{G}$ by $A$. It is important to note that the lift is not uniquely determined by the group $A$. \\

The theory of projective representations of finite groups over the complex number field was founded and developed by I. Schur \cite{Schur+1904+20+50,Schur+1907+85+137}. From Schur's lemma, it follows that the irreducible representations of central extensions of $G$, and the irreducible projective representations of $G$, are essentially the same objects. 
The reason why this is convenient is because we cannot take direct sums of projective representations unless multipliers agree. Basically, knowing the irreducible components of projective representations does not determine the representation itself. Therefore, it is convenient to work on a lift where we understand the ordinary representations.\\

In the case $n_g=1/2$, the group $\mathbb{Z}_2 \times \mathbb{Z}_2$ is lifted to $D_4$ by the abelian group $\mathbb{Z}_2$. Note that $\mathbb{Z}_2 \times \mathbb{Z}_2$ can also be lifted to the quaternion group $Q_8$ by $\mathbb{Z}_2$. The elements of the group $\tilde{G}= D_{4}$ are listed below,
\[
\{\pm \sigma^{0}, \pm \sigma^{x},\pm \sigma^{z},\pm \sigma^{z}\sigma^{x}\}; 
\]
and we get the following diagram:
\begin{equation}
\begin{split}
1 \xrightarrow ~ \mathbb{Z}_2 \xrightarrow{\iota} D_4 &\xrightarrow{\mathcal{\pi}} \mathbb{Z}_2 \times \mathbb{Z}_2 \xrightarrow ~ 1 \\
1 \xrightarrow ~ 1 \xrightarrow{\iota}  \sigma^{0} &\xrightarrow{\mathcal{\pi}} \bar{\sigma^{0}}
\xrightarrow ~ 1 \\
 -1 \xrightarrow{\iota} - \sigma^{0} &\xrightarrow{\mathcal{\pi}}  \sigma^{0}
\xrightarrow ~ 1 \\
\sigma^{x} & \xrightarrow{\mathcal{\pi}} \sigma^{x}
\xrightarrow ~ 1 \\
-\sigma^{x} & \xrightarrow{\mathcal{\pi}} \sigma^{x}
\xrightarrow ~ 1 \\
\sigma^{z} &\xrightarrow{\mathcal{\pi}} \sigma^{z}
\xrightarrow ~ 1\\
- \sigma^{z} &\xrightarrow{\mathcal{\pi}} \sigma^{z}
\xrightarrow ~ 1\\
\sigma^{z} \sigma^{x} &\xrightarrow{\mathcal{\pi}} \sigma^{z}\sigma^{x} 
\xrightarrow ~ 1\\
- \sigma^{z} \sigma^{x} &\xrightarrow{\mathcal{\pi}} \sigma^{z} \sigma^{x}
\xrightarrow ~ 1
\end{split}
\end{equation}

A different lift for $\mathbb{Z}_2\times \mathbb{Z}_2$ by $\mathbb{Z}_2$ is $Q_8$. In this case $L(\bar{\sigma^{x}})=i\sigma^{x}, L(\bar{\sigma^{z}})=i\sigma^{z}$ and consequently $L(\bar{\sigma^{z} \sigma^{x}})=i\sigma^{y}$. Hence identifying $a\to L(\sigma^{x})$ and $b\to L(\sigma^{z})$ we would obtain the following relations:
\[
a^4=1, a^2=b^2, bab^{-1}=a^{-1}
\]

\subsection{Perturbative SW}
\label{sec:pert-schr-wolff}

\subsubsection{General framework}
\label{sec:general-framework}  

The SW method is an approximation method that provides us with effective Hamiltonians by successive improvements on block diagonalisation. There are many presentations and variants of it since its original introduction in physics in 1966 \cite{SW1966}. Here we will mostly follow the one of Bravyi et al \cite{Bravyi2011}, adapted to the case at hand.

In all SW variants, the organisation of the successive approximations is achieved by introducing a $\mathbb{Z}_2$ grading in the Lie algebra of operators (we shall use a presentation and terminology adequate for finite-dimensional Hilbert spaces, even though we shall apply the method to an infinite-dimensional one). The \emph{even} operators will be those that are block-diagonal with respect to what is called a $PQ$ partitioning, while the \emph{odd} ones are block-off-diagonal. The partitioning is determined by a projector $P$ (the projector onto the low-energy sector, in most applications) and its complementary $Q=\mathbbm{1}-P$, in such a way that the even ($X_{\mathrm{e}}$) and odd  ($X_{\mathrm{o}}$) elements of an operator $X$ are
\begin{align}
\label{eq:SWevenodd}
X_{\mathrm{e}}&= PXP +QXQ\,,\\
X_{\mathrm{o}}&= PXQ +QXP\,.\nonumber
\end{align}
This is a grading of the Lie algebra in which the commutator of two even or two odd elements is even, while the commutator of one even and one odd element is odd.

The fundamental idea in the SW method is to identify successive approximations to an antihermitian odd operator $S$, which, if known in full, would make $e^{S}He^{-S}$  an even operator. For clarity we shall use super-operator notation. In particular, 
\begin{subequations}
\label{eq:superops}
\begin{align}
\mathrm{Ad}_X(Y)& = \left[X,Y\right]\label{eq:addef}\,,\\
\mathrm{E}(X)& = X_{\mathrm{e}}= PXP +QXQ\,,\label{eq:evendef}\\
\mathrm{O}(X)& = X_{\mathrm{o}}= PXQ +QXP\,.\label{eq:odddef}
\end{align}
\end{subequations}
Thus, formally, we require
\begin{equation}
\label{eq:SWcond}
\mathrm{O}\circ \exp\left[\mathrm{Ad}_S\right] \left(H\right)=0\,.
\end{equation}
In what follows we shall only explicitly write the composition symbol when it aids in reading a formula.

Observe that since $S$ is an odd operator, the expansion of $\exp\left[\mathrm{Ad}_S\right]$ when acting on even operators will have an even part determined by even powers, while the odd part will come from odd powers. Thus, formally,
\begin{subequations}
\label{eq:expdecom}
\begin{align*}
\mathrm{E}\circ \exp\left[\mathrm{Ad}_S\right]\circ \mathrm{E} &=   \mathrm{E}\circ \cosh\left[\mathrm{Ad}_S\right]\circ \mathrm{E}=\cosh\left[\mathrm{Ad}_S\right]\circ \mathrm{E} \,,\\
\mathrm{E}\circ \exp\left[\mathrm{Ad}_S\right]\circ \mathrm{O} &=   \mathrm{E}\circ \sinh\left[\mathrm{Ad}_S\right]\circ \mathrm{O}=\sinh\left[\mathrm{Ad}_S\right]\circ \mathrm{O}\,,\\
\mathrm{O}\circ \exp\left[\mathrm{Ad}_S\right]\circ \mathrm{E} &=   \mathrm{O}\circ \sinh\left[\mathrm{Ad}_S\right]\circ \mathrm{E}=\sinh\left[\mathrm{Ad}_S\right]\circ \mathrm{E}\,,\\
\mathrm{O}\circ \exp\left[\mathrm{Ad}_S\right]\circ \mathrm{O} &=   \mathrm{O}\circ \cosh\left[\mathrm{Ad}_S\right]\circ \mathrm{O}=\cosh\left[\mathrm{Ad}_S\right]\circ \mathrm{O}\,,
\end{align*}
\end{subequations}
exactly as parity extracts the even and odd parts of an exponential function.

Therefore, the SW condition $\mathrm{O}\circ\exp\left[\mathrm{Ad}_S\right](H)=0$ can be written as
\begin{equation}
\label{eq:rewriteSW}
\sinh\left[\mathrm{Ad}_S\right]\circ\mathrm{E}(H)+
\cosh\left[\mathrm{Ad}_S\right]\circ\mathrm{O}(H)=0\,.
\end{equation}
An alternative formal rewriting of this condition that will prove useful reads
\begin{equation}
\label{eq:secondrewriteSW}
\mathrm{Ad}_S\circ E(H)+\mathrm{Ad}_S\coth\left[\mathrm{Ad}_S\right]\circ\mathrm{O}(H)=0\,.
\end{equation}

If this condition is met, then the effective Hamiltonian is
\begin{equation}
\label{eq:effec}
H_{\mathrm{eff}}= E(H)+\tanh\left[\frac{1}{2}\mathrm{Ad}_S\right]\circ\mathrm{O}(H)\,.
\end{equation}

These formal expressions are, by themselves, not terribly useful. They have to be implemented by successive approximations, determined by a formal expansion $S=\sum_{n=1}^\infty S_n$ and a criterion for selecting orders of the expansion. Frequently this is achieved by considering the Hamiltonian as composed of two parts,
\begin{equation}
\label{eq:hamexp}
H= H_0+V= \left(H_0+V_{\mathrm{e}}\right)+ V_{\mathrm{o}}\,,
\end{equation}
and performing the formal expansion in terms of powers of $V$, where both its even and odd parts are of the same formal order. In eq. \eqref{eq:hamexp} the assumption that $H_0$ is even has also been made explicit, that is $\left[P,H_0\right]=0$. Although the method can be tweaked to the degenerate case, let us consider that $H_0$ has no degenerate eigenenergies. Acting on odd operators a formal inverse of $\mathrm{Ad}_{{H_0}}$, that we denote as $\mathrm{L}$, exists,
\begin{equation}
\label{eq:inverse}
\mathrm{L}\circ\mathrm{Ad}_{{H_0}}\circ\mathrm{O}=\mathrm{O}\,,
\end{equation}
given by
\begin{equation}
\label{eq:lx}
\mathrm{L}\left[\mathrm{O}(X)\right]=\sum_{N,M}\left|N\right\rangle\frac{\left\langle N\right|\mathrm{O}(X)\left| M\right\rangle}{E_N-E_M}\left\langle M\right|\,,
\end{equation}
with $|N\rangle$ and $|M\rangle$ running over the eigenstates of $H_0$ with energies $E_N$ and $E_M$ respectively.

Since $S$ is odd, eq. \eqref{eq:inverse} entails
\begin{equation}
\label{eq:ash0}
\mathrm{L}\circ\mathrm{Ad}_S\left(H_0\right)= -S\,.
\end{equation}
Applying the formal inverse $\mathrm{L}$ to the SW condition in the form of eq. \eqref{eq:secondrewriteSW} one obtains
\begin{equation}
\label{eq:Sformal}
S= \mathrm{L}\circ\mathrm{Ad}_S\left(V_{\mathrm{e}}\right)+\mathrm{L}\circ\mathrm{Ad}_S\coth\left[\mathrm{Ad}_S\right]\left(V_{\mathrm{o}}\right)\,.
\end{equation}
Observe that the formal inverse is always applied to odd operators in this expression. Furthermore, it is now amenable to perturbative treatment. Namely, assume that $S_n$ is of formal order $V^n$, and identify orders in both sides. In this way we have a determination of $S_n$ from $V$ and $S_m$ with $m<n$. The first terms of the expansion are
\begin{subequations}
\label{eq:BdVLexpansion}
\begin{align}
S_1&= \mathrm{L}\left(V_{\mathrm{o}}\right)\,,\label{eq:BdVL1}\\
S_2&= \mathrm{L}\circ\mathrm{Ad}_1\left( V_{\mathrm{e}}\right)\,,\label{eq:BdVL2}\\
S_3&= \mathrm{L}\circ\left[\mathrm{Ad}_2 \left(V_{\mathrm{e}}\right)+ \frac{1}{3} \mathrm{Ad}_1^2\left(V_{\mathrm{o}}\right)\right]\,.\label{eq:BdVL3}
\end{align}
\end{subequations}
$\mathrm{Ad}_n$ has been used to denote
\begin{equation}
\label{eq:adndef}
\mathrm{Ad}_n\left(X\right)= \left[S_n,X\right]\,.
\end{equation}
Correspondingly, the first terms of the effective Hamiltonian are \begin{align}
\label{eq:effsol}
H_{\mathrm{eff},\mathrm{exact}}&= H_{\mathrm{e}}+3 B_2 \mathrm{Ad}_1V_{\mathrm{o}}+3 B_2 \mathrm{Ad}_2V_{\mathrm{o}}+O\left(V^4\right)\\
&= H_{\mathrm{e}}+\frac{1}{2} \mathrm{Ad}_1V_{\mathrm{o}}+\frac{1}{2} \mathrm{Ad}_2V_{\mathrm{o}}+O\left(V^4\right)\,.
\end{align}

\subsubsection{Symmetries}
\label{sec:symmetries}

In the case we are studying the Hilbert space is factorised, $\mathcal{H}=\mathcal{H}_\phi\otimes\mathcal{H}_\theta$. The projector being used is actually of the form $P_\phi\otimes\mathbbm{1}$. Given any unitary $U$ it induces the super-operator $\mathrm{U}$ by conjugation,
\begin{equation}
\label{eq:Usuper}
\mathrm{U}\left(X\right)= U XU^\dag\,.
\end{equation}
Its composition with $\mathrm{Ad}_X$ obeys the rule
\begin{equation}
\label{eq:uad}
\mathrm{U}\circ\mathrm{Ad}_X=\mathrm{Ad}_{\mathrm{U}(X)}\circ\mathrm{U}\,.
\end{equation}

Consider now a unitary acting only on the second subspace, $\mathbbm{1}\otimes U$. Its corresponding super-operator $\mathrm{U}$ commutes with super-operators $\mathrm{E}$, $\mathrm{O}$ and $\mathrm{L}$, and leaves $H_0$ invariant, $U\left(H_0\right)=H_0$.

If this unitary is a symmetry of the Hamiltonian, it must be a symmetry of $V$, and furthermore it is separately a symmetry of its even and odd parts, $\mathrm{U}\left(V_{\mathrm{e}}\right)=V_{\mathrm{e}}$ and $\mathrm{U}\left(V_{\mathrm{o}}\right)=V_{\mathrm{o}}$. Therefore, applying $\mathrm{U}$ on both sides of eq. \eqref{eq:Sformal} one determines that if $S$ is a solution so is $\mathrm{U}(S)$. Furthermore, applying $\mathrm{U}$ to the perturbative expansion \eqref{eq:BdVLexpansion} one determines that it is invariant order by order under this symmetry, i.e., $\mathrm{U}\left(S_n\right)=S_n$.

There is another type of symmetry to be examined. Let the symmetry be realised by a unitary that factorises as $U=U_\phi\otimes U_\theta$, where $U_\phi$ is an involution and a symmetry of $H_0$, while $U$ is a symmetry of $H$. In such a  case $U_\phi|N\rangle\langle N|=|N\rangle\langle N|U_\phi$, and it follows that $\mathrm{U}$ commutes with $\mathrm{L}$. Additionally, $U_\phi P = P U_\phi$. Furthermore, $U$ also commutes with $V_{\mathrm{e}}$ and  $V_{\mathrm{o}}$ separately. It again follows that  $\mathrm{U}\left(S_n\right)=S_n$ for this type of symmetry.

Applying these two results to the case under study, we see that our perturbative SW expansion respects the exceptional symmetry, if present, order by order.

\subsection{First- and second-order expansion in the SW operator}

The first step in the SW transformation is to split the total Hilbert space of the model in a ``low-energy'' part, described by the projector $| 0_{\phi} \rangle \langle 0_{\phi} | \equiv  P_{\phi}$ in our case, and the ``high-energy''  part, $ \sum_{N_{\phi}\neq 0} |N_{\phi} \rangle \langle N_{\phi}| \equiv  Q_{\phi}$, with $\mathbb{I}_{\phi} = P_{\phi} + Q_{\phi}$. It is easy to check that both projectors commute with the all symmetry operators, i.e., this splitting respects the symmetry of the model.

With this splitting, the $H_{0\text{-}\pi}$ Hamiltonian is also reorganised in terms that keep the dynamics within the ``low-'' or ``high-energy'' sectors and the ones that connect them, i.e.,
\begin{equation}
H_{0\text{-}\pi} =H_{0} + V_{\text{even}} + V_{\text{odd}},
\end{equation}
with $H_{0} =2 \sqrt{E_{C_J} E_L} \left( \hat{N}_{\phi} + 1/2 \right) $, with
\begin{equation}
\begin{split}
V_{\text{even}} =& P_{\phi} V P_{\phi} + Q_{\phi} V Q_{\phi}\\
=&|0_{\phi}\rangle c_{0,0} \langle 0_{\phi}| + \sum_{N_{\phi},M_{\phi}\neq 0} |N_{\phi} \rangle c_{N,M} \langle M_{\phi}|\\
V_{\text{odd}} =& P_{\phi} V Q_{\phi} + Q_{\phi} V P_{\phi} \\
=& \sum_{N_{\phi} \neq 0} |N_{\phi} \rangle c_{N,0} \langle 0_{\phi}|+  |0_{\phi} \rangle c_{0,N} \langle N_{\phi}|,
\end{split}
\end{equation}
and $c_{N,M}= \langle  N_{\phi} | V | M_{\phi} \rangle$.

At this point, SW is an approximation method that provides us with effective Hamiltonians by successive improvements on block diagonalisation, reducing order by order the terms in the Hamiltonian that connect ``low-'' and ``high-energy'' sectors. Formally, there is a unitary transformation generated by the SW operator $S$ such that
\begin{equation}
\begin{split}
&\tilde{H}_{0\text{-}\pi} = e^{S} H_{0\text{-}\pi} e^{-S} = e^{ \left[ S , \right] } H_{0\text{-}\pi} \equiv \\
& H_{0\text{-}\pi} +  [S , H_{0\text{-}\pi} ] + \frac{1}{2 !} \left[ S, \left[ S , H_{0\text{-}\pi} \right] \right] + \cdots ,
\end{split}
\end{equation}
where the exponentiated commutator is defined by the series expansion on the second line. We decomposed the operator $S=\sum_{n=1} S_{n}$ as a sum of $S_{n}$ operators that should eliminate order by order the off-diagonal elements of the effective Hamiltonian. 

For instance, the first order expansion of the exponentiated commutator fixes $V_{\text{odd}} + \left[ S_{1} , H_{0}\right]=0$, the second order $ \left[ S_{1} , V_{\text{even}}\right]+ \left[ S_{2} , H_{0}\right]=0$, which give
\begin{equation*}
\begin{split}
S_{1} =& \sum_{N_{\phi} \neq 0} \frac{c_{N,0}}{2 \sqrt{E_{C_J} E_L} N} |N_{\phi} \rangle  \langle 0_{\phi}| - \frac{c_{0,N}}{2 \sqrt{E_{C_J} E_L} N} |0_{\phi} \rangle  \langle N_{\phi}|, \\
S_{2} =& \sum_{N_{\phi} \neq 0} \left( \frac{c_{N,0} c_{0,0} }{4 E_{C_J} E_L N^{2}} -  \sum_{M_{\phi} \neq 0} \frac{c_{N,M}c_{M,0}}{4 E_{C_J} E_L NM} \right) |N_{\phi} \rangle  \langle 0_{\phi}| \\
&- \left( \frac{c_{0,0}c_{0,N}}{4 E_{C_J} E_L N^{2}} -  \sum_{M_{\phi} \neq 0} \frac{c_{0,M}c_{M,N}}{4 E_{C_J} E_L NM}  \right) |0_{\phi} \rangle  \langle N_{\phi}|.
\end{split}
\end{equation*} 

Given that all the operators that define the SW generator keep the symmetry of the model, this is guaranteed also in the definition order by order of the generator $S$. Up to this second order, the effective Hamiltonian is of the form
\begin{equation}
\tilde{H}_{0-\pi} = H_{0} + V_{\text{even}} + \frac{1}{2} \left[ S_{1}, V_{\text{odd}} \right] + \frac{1}{2} \left[ S_{2}, V_{\text{odd}} \right] 
\end{equation}
This expression is the standard form of the SW transformation. Note that all the operators on the right-hand side are now expressed in a new basis \emph{dressed} by the interaction to second order. The Hamiltonian can be projected on the lowest subspace to obtain an effective projected Hamiltonian for low-energy dynamics. In order for the transformation to be accurate, the eliminated subspaces must be energetically well separated from the subspace of interest, meaning that the strength of the interaction $E_{J}$ must be much smaller than the energy difference between the subspaces $\sqrt{E_{C_J} E_L}$. 

\subsection{Matrix elements for the SW expansion}
\label{sec:matrix-elements-sw}
We need the matrix elements $\langle M_{\phi} | D\left(\alpha \right) | N_{\phi} \rangle$, where $|N_{\phi} \rangle$, $|M_{\phi} \rangle$ are Fock states with $N$ and $M$ excitations, and $D\left(\alpha \right) = e^{\alpha a^{\dagger}_{\phi} - \alpha^{*} a_{\phi}} = e^{\alpha a^{\dagger}_{\phi}} e^{-|\alpha|^{2}/2} e^{-\alpha^{*}a_{\phi}}$ is the displacement operator. Knowing that
\begin{equation}
e^{-\alpha^{*}a_{\phi}} | N_{\phi} \rangle = \sum^{N}_{K_{\phi}=0} \frac{(-\alpha^{*})^{N-K}}{(N-K)!} \sqrt{\frac{N!}{K!}} |K_{\phi} \rangle
\end{equation}
then, it follows  
\begin{equation}
\begin{split}
&\langle M_{\phi} | D\left(\alpha \right) | N_{\phi} \rangle = \langle M_{\phi} | e^{\alpha a_{\phi}^{\dagger}} e^{-|\alpha|^{2}/2} e^{-\alpha^{*}a_{\phi}} | N_{\phi} \rangle \\
&= e^{-|\alpha|^{2}/2} \sqrt{M!N!} \sum^{\min(M,N)}_{K_{\phi}=0}  \frac{(\alpha)^{M-K}(-\alpha^{*})^{N-K}}{(M-K)!(N-K)! K!} 
\end{split}
\end{equation}
From this equation, the next matrix elements follow
\begin{equation*}
\begin{split}
&c_{M,N}= \langle  M_{\phi} | V | N_{\phi} \rangle = \\
=&\langle  M_{\phi} | 4E_{C_s} \left( \hat{n}_{\theta} - n_{g} \right)^2 -2E_J\cos \theta\cos \left( \phi-\frac{\varphi_{\mathrm{ext}}}{2} \right) | N_{\phi} \rangle \\
=&4E_{C_s} \left( \hat{n}_{\theta} - n_{g} \right)^2 \delta_{NM} \\
&- \frac{2E_J\cos \theta e^{ - \frac{1}{2}\sqrt{ \frac{E_{C_J}}{E_L} }}  i^{M+N}}{\sqrt{N! M!}} \left(\frac{E_{C_J}}{E_L} \right)^{\frac{N-M}{4}} \\
& \cdot \mathcal{U} \left[ -M , 1-M+N, \sqrt{\frac{E_{C_J} }{ E_L}} \right] \\
&\cdot \left\{ \begin{array}{rl} \cos\left( \frac{\varphi_{\mathrm{ext}}}{2}\right)& M+N \in \text{even} \\ (-i)\sin\left( \frac{\varphi_{\mathrm{ext}}}{2}\right)& M+N \in \text{odd} \end{array}\right.
\end{split}
\end{equation*}
where $\mathcal{U} \left[ -M , 1-M+N, \sqrt{\frac{E_{C_J} }{ E_L}} \right]$ is the confluent hypergeometric function. From this equation, there are some exact and explicit expressions that we will also show in the asymptotic limit $E_{C_J} \gg E_L$. For instance

\begin{equation*}
\begin{split}
&\frac{1}{2}\langle 0_{\phi} | \left[ S_{1} ,V_{\text{odd}} \right]| 0_{\phi} \rangle = - \sum_{N_{\phi} \neq 0} \frac{c_{0,N}c_{N,0}}{2 \sqrt{E_{C_J} E_L} N} \\
=& \frac{- 2 E_{J}^{2} \cos^{2} \theta e^{ - \sqrt{ \frac{E_{C_J}}{E_L} }}  }{ \sqrt{E_{C_J} E_L} } \Bigg\{ \\
& \left[ - \gamma_{E} + \text{Chi}\left( \sqrt{ \frac{E_{C_J} }{E_L} } \right)  - \log{\left( \sqrt{ \frac{E_{C_J} }{E_L} } \right)} \right] \cos^{2} \left( \frac{\varphi_{\mathrm{ext}}}{2} \right) \\
&+ \text{Shi}\left( \sqrt{ \frac{E_{C_J} }{E_L} } \right)  \sin^{2} \left( \frac{\varphi_{\mathrm{ext}}}{2} \right) \Bigg\}
\end{split}
\end{equation*}
where $\text{Chi}$ (\text{Shi}) is the hyperbolic cosine integral (hyperbolic sine integral).

For the  third-order correction to the SW expansion, we get, in the effective Hamiltonian, the term
\begin{equation*}
\begin{split}
&\frac{1}{2} \langle 0_{\phi} | \left[ S_{2} ,V_{\text{odd}} \right]| 0_{\phi} \rangle =  \sum_{N_{\phi} \neq 0} \\
& \left[ \sum_{M_{\phi} \neq 0} \left( \frac{ c_{0,N} c_{N,M} c_{M,0}}{ 4 E_{C_J} E_L NM} \right) - \frac{c_{0,N} c_{N,0} c_{0,0} +c_{0,0} c_{0,N} c_{N,0} }{8 E_{C_J} E_L N^{2}} \right]
\end{split} 
\end{equation*}

Knowing that
\begin{equation*}
\begin{split}
&\sum_{N_{\phi} \neq 0} \frac{c_{0,N} c_{N,0}}{N^{2}} = 4 E^{2}_{J} \cos^{2}\left( \theta \right) e^{ -\sqrt{ \frac{E_{C_J}}{E_L} }} \\
&\Bigg\{ \frac{E_{C_J}}{8 E_{L}} \cos^{2}\left(\frac{\varphi_{\mathrm{ext}}}{2}\right) ~ _{3}F_{4} \left[ \{ 1,1,1 \} , \{ 3/2 , 2 , 2, 2 \} , \frac{E_{C_J}}{4 E_{L}} \right] + \\
&\sqrt{\frac{E_{C_J}}{E_{L}}} \sin^{2}\left(\frac{\varphi_{\mathrm{ext}}}{2}\right) ~ _{2}F_{3} \left[ \{ 1/2 ,1/2 \} , \{ 3/2 , 3/2 , 3/2 \} , \frac{E_{C_J}}{4 E_{L}} \right] \Bigg\} \\
&\xrightarrow[ E_{C_{J}} \gg E_{L} ] ~ 4 E^{2}_{J} \cos^{2}\left( \theta \right) e^{ - \sqrt{ \frac{E_{C_J}}{E_L} }} \\
&\Bigg\{  \frac{e^{ \sqrt{ \frac{E_{C_J}}{E_L} }} }{2 \left(  E_{C_J} / E_{L}  \right)^{3/2} }  \left[  3 \sin^{2}\left(\frac{\varphi_{\mathrm{ext}}}{2}\right) +  \sqrt{\frac{ E_{C_J}  }{E_{L} } }   \right] \Bigg\}
\end{split}
\end{equation*}
with $_{p}F_{q} \left[ a ; b  ; z \right]$ the generalised hypergeometric function, we get an expression for the third-order correction for the terms proportional to $\sum \frac{c_{0,N} c_{N,0}}{N^{2}} $
\begin{equation*}
\begin{split}
& \Bigg\{ \frac{2 E_{J}^{2} E_{C_{s}} }{ E_{C_{J}} E_{L}} e^{ - \sqrt{ \frac{E_{C_J}}{E_L} }} \sin^{2}\left( \theta \right) \\
&+ \frac{2 E_{J}^{3}  \cos{ \left( \varphi_{\mathrm{ext}} / 2 \right) }   }{E_{C_{J}} E_{L}}  e^{ - \frac{3}{2}\sqrt{ \frac{E_{C_J}}{E_L} }} \cos^{3}\left( \theta \right) \Bigg\} \\
&\Bigg\{ \frac{E_{C_J}}{8 E_{L}} \cos^{2}\left(\frac{\varphi_{\mathrm{ext}}}{2}\right) ~ _{3}F_{4} \left[ \{ 1,1,1 \} , \{ \frac{3}{2} , 2 , 2, 2 \} , \frac{E_{C_J}}{4 E_{L}} \right] + \\
&\sqrt{\frac{E_{C_J}}{E_{L}}} \sin^{2}\left(\frac{\varphi_{\mathrm{ext}}}{2}\right) ~ _{2}F_{3} \left[ \{ 1/2 ,1/2 \} , \{ \frac{3}{2} , 3/2 , 3/2 \} , \frac{E_{C_J}}{4 E_{L}} \right] \Bigg\}  \\
& \xrightarrow[ E_{C_{J}} \gg E_{L} ] ~  \Bigg\{ \frac{2 E_{J}^{2} E_{C_{s}} }{ E_{C_{J}} E_{L}} e^{ -\sqrt{ \frac{E_{C_J}}{E_L} }} \sin^{2}\left( \theta \right) \\
&+ \frac{2 E_{J}^{3}   }{E_{C_{J}} E_{L}}  e^{ - \frac{3}{2}\sqrt{ \frac{E_{C_J}}{E_L} }} \cos{ \left( \frac{\varphi_{\mathrm{ext}}}{2}  \right) } \cos^{3}\left( \theta \right) \Bigg\} \\
&\Bigg\{  \frac{ e^{ \sqrt{ \frac{E_{C_J}}{E_L} }} }{2 \left(  E_{C_J} / E_{L}  \right) }  \left[  3 ~ \sqrt{ \frac{E_{L} } { E_{C_J} } } \sin^{2}\left(\frac{\varphi_{\mathrm{ext}}}{2}\right) +  1   \right] \Bigg\} = \\
&\frac{E_{J}^{2} E_{C_{s}} }{ E^{2}_{C_{J}} }  \sin^{2}\left( \theta \right) \left[  3 ~ \sqrt{ \frac{E_{L} } { E_{C_J} } } \sin^{2}\left(\frac{\varphi_{\mathrm{ext}}}{2}\right) +  1   \right] \\
&+ \frac{ E_{J}^{3}    }{E^{2}_{C_{J}} }  e^{ - \frac{1}{2}\sqrt{ \frac{E_{C_J}}{E_L} }} \cos{ \left( \frac{\varphi_{\mathrm{ext}}}{2}  \right) } \cos^{3}\left( \theta \right)
\end{split}
\end{equation*}

For the term that involved the sum with $\mathcal{U}$ factor, we can give a bound knowing that $\mathcal{U} \left[ -M , 1-M+N, \sqrt{\frac{E_{C_J} }{ E_L}} \right] \xrightarrow[ E_{C_{J}} \gg E_{L} ] ~ \left(\frac{ E_{C_{J}} }{ E_{L} }\right)^{M/2} \left( 1 - MN \sqrt { \frac{ E_{L} }{ E_{C_{J}}  } }\right)$, then, this term gives
\begin{equation*}
\frac{ E^{3}_{J} }{ E_{C_{J}} \sqrt{ E_{L} E_{C_{J}} } } \log{\left(\frac{E_{C_{J}}}{E_{L}} \right)}  e^{-\frac{1}{2} \sqrt{\frac{E_{C_{J}}}{E_{L}}}}\cos{ \left( \frac{\varphi_{\mathrm{ext}}}{2}  \right) }   \cos^{3}\left( \theta \right) 
\end{equation*}

Summing the different parts gives
\begin{equation*}
\begin{split}
&\frac{1}{2} \langle 0_{\phi} | \left[ S_{2} ,V_{\text{odd}} \right]| 0_{\phi} \rangle  \xrightarrow[ E_{C_{J}} \gg E_{L} ] ~ \\
&\frac{E_{J}^{2} E_{C_{s}} }{ E^{2}_{C_{J}} }  \sin^{2}\left( \theta \right) \left[  3 ~ \sqrt{ \frac{E_{L} } { E_{C_J} } } \sin^{2}\left(\frac{\varphi_{\mathrm{ext}}}{2}\right) +  1   \right] \\
&+ \frac{ E_{J}^{3}    }{E^{2}_{C_{J}} }  e^{ - \frac{1}{2}\sqrt{ \frac{E_{C_J}}{E_L} }} \cos{ \left( \frac{\varphi_{\mathrm{ext}}}{2}  \right) } \cos^{3}\left( \theta \right) \\
&+ \frac{ E^{3}_{J} }{ E_{C_{J}} \sqrt{ E_{L} E_{C_{J}} } } \log{\left(\frac{E_{C_{J}}}{E_{L}} \right)}  e^{-\frac{1}{2} \sqrt{\frac{E_{C_{J}}}{E_{L}}}}\cos{ \left( \frac{\varphi_{\mathrm{ext}}}{2}  \right) }   \cos^{3}\left( \theta \right) 
\end{split} 
\end{equation*}

As a summary, explicitly breaking the initial symmetry of the model, i.e., at $(n_{g},\varphi_{\mathrm{ext}}) \neq (1/2,\pi)$, we get for the $H_{0\text{-}\pi}$ model
\begin{equation*}
\begin{split}
& \langle 0_{\phi} | V_{\text{even}} | 0_{\phi} \rangle = 4E_{C_s} \left( \hat{n}_{\theta} - n_{g} \right)^2 \\
&-2E_J\cos \theta \cos \left( \frac{\varphi_{\mathrm{ext}}}{2} \right) e^{ - \frac{1}{2}\sqrt{ \frac{E_{C_J}}{E_L} }}.\\
&\frac{1}{2}\langle 0_{\phi} | \left[ S_{1} ,V_{\text{odd}} \right]| 0_{\phi} \rangle \xrightarrow[ E_{C_{J}} \gg E_{L} ] ~ \frac{-E_{J}^{2}\cos^2{\theta}}{  E_{C_J} }   \\
&  \left[ 1 + \sqrt{\frac{E_{L}}{E_{C_J}}} - e^{-\sqrt{ \frac{E_{C_J}}{E_L}}} \sqrt{ \frac{E_{C_J}}{E_L}}  \log{ \left( \frac{E_{C_J}}{E_L} \right)} \cos^{2} \left( \frac{\varphi_{\mathrm{ext}}}{2} \right) \right]. \\
&\frac{1}{2} \langle 0_{\phi} | \left[ S_{2} ,V_{\text{odd}} \right]| 0_{\phi} \rangle \xrightarrow[ E_{C_{J}} \gg E_{L} ] ~ \\
& \frac{E_{J}^{2} E_{C_{s}} }{ E^{2}_{C_{J}} }  \sin^{2}\left( \theta \right) \left[  3 ~ \sqrt{ \frac{E_{L} } { E_{C_J} } } \sin^{2}\left(\frac{\varphi_{\mathrm{ext}}}{2}\right) +  1   \right] \\
&+ \frac{ E_{J}^{3}  e^{-\frac{1}{2} \sqrt{\frac{E_{C_{J}}}{E_{L}}}}\cos{ \left( \frac{\varphi_{\mathrm{ext}}}{2}  \right) }   \cos^{3}\left( \theta \right)  }{E_{C_{J}} }   \left[ \frac{ 1   }{E_{C_{J}} }+ \frac{ \log{\left(\frac{E_{C_{J}}}{E_{L}} \right)}  }{  \sqrt{ E_{L} E_{C_{J}} } }  \right] .
\end{split}
\end{equation*}

\subsection{Charge offset sensitivity}
\label{sec:charge-offs-sens}

Let us consider Hamiltonians of the form
\begin{equation}
  \label{eq:gen2theta}
  H= 4E_{C_{s}} \left(\hat{n}_\theta-n_g\right)^2-\sum_{k\in\mathbb{N}}\lambda_k\cos(2k\theta)\,.
\end{equation}
They all present a discrete translation symmetry, $\hat{U}_\pi=e^{i\pi\hat{n}_\theta}$, which is an involution and maps $|n\rangle\to(-1)^n|n\rangle$. As a consequence, the spectra are classified in  even and odd sectors, with only even charges ($|2n\rangle$) in the even and odd charges ($|2n+1\rangle$) in the odd sector. The involution $\hat{V}_P$ maps $|n\rangle$ to $|1-n\rangle$, and therefore maps the even to the odd sector.

For a finite number of cosine terms, the potential is a bounded operator. Its effect on the eigenenergies, therefore, becomes negligible for high occupancies, so the eigenvectors for high values of the energy are close to number vectors, $|n\rangle$. The corresponding energies are not very sensitive to changes in the charge  offset, therefore. The real interest, though, lies in the lowest part of the spectrum. In fact, as we know that at the special $n_g=1/2$ symmetry point the ground level is degenerate, we characterise the sensitivity by the opening of an energy gap at that point. Observe that for the free case ($\lambda_k$ all equal to 0) the gap as a function of $n_g$ is readily computed,
\begin{equation}
  \label{eq:freegap}
  |\Delta(\mathrm{free})|=\left|E_1-E_0 \right|=4E_{C_{s}}\left|1-2n_g\right|\,.
\end{equation}
Here the lowest lying even state is $|n=0\rangle$, while the lowest lying odd state is $|n=1\rangle$. We shall define thus
\begin{equation}
  \label{eq:deltadef}
  \Delta=E(\mathrm{gs},e)-E(\mathrm{gs},o)\,,
\end{equation}
where $E(\mathrm{gs},o)$ (resp. $E(\mathrm{gs},e)$) is the ground-state energy in the odd (resp. even) sector, and the magnitude characterising sensitivity will be $\delta(n_g)=\partial\Delta/\partial n_g$. At $n_g=1/2$, using $|\mathrm{gs},o\rangle=\hat{V}_P|\mathrm{gs},e\rangle$ at that point and the Feynman--Hellmann theorem, one observes
\begin{align}
  \label{eq:sensitivitygeneral}
  \delta(1/2)&= \frac{\partial E(\mathrm{gs},e)}{\partial n_g}- \frac{\partial E(\mathrm{gs},o)}{\partial n_g}\nonumber\\
             &=\left\langle\mathrm{gs},e\Big|\frac{\partial H}{\partial n_g}\Big|\mathrm{gs},e\right\rangle-\left\langle\mathrm{gs},o\Big|\frac{\partial H}{\partial n_g}\Big|\mathrm{gs},o\right\rangle\nonumber\\
             &=\left\langle\mathrm{gs},e\Big|\left[\frac{\partial H}{\partial n_g}-\hat{V}_P\frac{\partial H}{\partial n_g}\hat{V}_P\right]\Big|\mathrm{gs},e\right\rangle\\
             &= -8E_{C_{s}}\left\langle\mathrm{gs},e\big|\left(\hat{n}_\theta-\hat{V}_P\hat{n}_\theta\hat{V}_P\right)\big|\mathrm{gs},e\right\rangle\nonumber\\
  &= 16 E_{C_{s}}\left(\frac{1}{2}-\left\langle\mathrm{gs},e|\hat{n}_\theta|\mathrm{gs},e\right\rangle\right)\,.\nonumber
\end{align}
As we see, we have to compute the average charge (without offset) in a ground-state. From a simple perturbative computation for the cosine case, one sees that this average charge is displaced from its zero value for the free even ground-state.

We have indicated above that the transmon regime  \cite{{Koch2007}} for the $\cos(2\theta)$ case shows exponential decrease of tunnelling from 0 to $\pi$, and therefore of splitting, as $E_{C_{s}}/\lambda\to0$. This can be shown explicitly in several different ways. For $\lambda \gg E_{C_{s}}$, the ground-states are the positive weight superposition of even or odd number of charges respectively, i.e., 
\begin{equation}
\begin{split}
&|\text{gs}_{o}\rangle \to \left(2Z\right)^{-1/2} \sum_{n} e^{ -\sqrt{\frac{2 E_{C_{s}}}{\lambda}} \frac{\left( 2n +1 - n_{g} \right)^{2}}{2}  } |2n+1\rangle \\
&|\text{gs}_{e}\rangle \to  \left(2Z\right)^{-1/2} \sum_{n} e^{ -\sqrt{\frac{2 E_{C_{s}}}{\lambda}} \frac{\left( 2n - n_{g} \right)^{2}}{2}  } |2n\rangle 
\end{split}
\end{equation}
with $ Z= \sqrt{\pi} \left[\frac{\lambda}{2 E_{C_{s}}} \right]^{1/4} $. In this limit, using the Poisson summation formula $K^{1/4} \sum_{n} e^{-\pi K n^{2}} = K^{-1/4} \sum_{m} e^{-\pi m^{2} / K} $\cite{Kadanoff2000}, we obtain
\begin{equation}
\langle \text{gs}_{e} | \left( \hat{n}_{\theta} - 1/2 \right) |\text{gs}_{e}\rangle = -\frac{\pi}{4} \sqrt{ \frac{\lambda}{2 E_{C_{s}}} } e^{-\frac{\pi^{2}}{4} \sqrt{\frac{\lambda}{2E_{C_{s}}} }}.
\end{equation}

\subsection{Explicit breaking from unbalance}
\label{sec:explicit-break}

In the main text we have not delved on the case of the circuit parameters  not being pairwise identical, respectively for  capacitors, inductors and Josephson junctions. In the case of circuit \emph{balance}, i.e. those pairs presenting identical values, there are two modes that decouple from the ones we study, the center of mass one and a collective harmonic oscillator.  Were any deviation from this balanced scenario to occur, as is bound to happen in any physical implemententation of  circuit, the collective harmonic oscillator would couple to the modes we have studied. We now address the impact of this coupling. The useful Lagrangian that corresponds to Fig. \ref{fig: 0-pi circuit} if there is no circuit balance  is given by
\begin{equation*}
\begin{split}
\mathcal{L} = &\frac{C_{24}}{2} \left( \dot{\phi}_{2} - \dot{\phi}_{4}  \right)^{2} + \frac{C_{13}}{2} \left( \dot{\phi}_{1} - \dot{\phi}_{3}  \right)^{2} \\
+ & \frac{C^{J}_{12}}{2} \left( \dot{\phi}_{1} - \dot{\phi}_{2}  \right)^{2}  +\frac{C^{J}_{34}}{2} \left( \dot{\phi}_{3} - \dot{\phi}_{4}  \right)^{2}\\
+ E^{J}_{12} &\cos{\left( \phi_{1} - \phi_{2} + \frac{\varphi_{ext}}{2}\right)} + E^{J}_{34} \cos{\left( \phi_{3} - \phi_{4} - \frac{\varphi_{ext}}{2} \right)} \\
- & \frac{1}{2L_{23}} \left( \phi_{2} - \phi_{3} \right)^{2} -  \frac{1}{2L_{14}} \left( \phi_{1} - \phi_{4} \right)^{2}
\end{split}
\end{equation*}
To assess the effect of small deviations from circuit balance we rewrite the parameters in terms of their circuit balance means and deviations from it,  
\begin{equation}
\begin{split}
C_{24} = \frac{C}{2} \left( 1 - dC \right);& ~ C_{13} = \frac{C}{2} \left( 1 + dC \right); \\
C^{J}_{12} = \frac{C_{J}}{2} \left( 1 - dC_{J} \right);& ~ C^{J}_{34} = \frac{C_{J}}{2} \left( 1 + dC_{J} \right); \\
E^{J}_{12} = E_{J} \left( 1 - \frac{dE_{J}}{2} \right);& ~ E^{J}_{34} = E_{J} \left( 1 + \frac{dE_{J}}{2} \right); \\
\frac{1}{L_{23}} =\frac{1}{2L}  \left( 1-dL\right) ;& ~\frac{1}{L_{14}} = \frac{1}{2L} \left( 1+dL\right).
\end{split}
\end{equation}

As is well known, the  relevant mode variables are connected to the initial ones by 
\begin{equation}
\begin{split}
\phi &= \frac{1}{2} \left( \phi_{4} - \phi_{3} + \phi_{2} - \phi_{1} \right); \\
\xi &= \frac{1}{2} \left( -\phi_{4} - \phi_{3} + \phi_{2} + \phi_{1} \right); \\
\theta &= \frac{1}{2} \left( -\phi_{4} + \phi_{3} + \phi_{2} - \phi_{1} \right); \\
\Sigma &= \frac{1}{2} \left( \phi_{4} + \phi_{3} + \phi_{2} + \phi_{1} \right)\,.
\end{split}
\end{equation}
The Lagrangian in terms of these variables reads
\begin{equation}\label{eq:lagnewvars}
\begin{split}
\mathcal{L} = & \frac{C+C_{J}}{2} \dot{\theta}^{2} + \frac{C}{2} \dot{\xi}^{2} + \frac{C_{J}}{2} \dot{\phi}^{2} \\
+ &2 E_{J} \cos{\left(\theta \right)} \cos{\left(\phi - \frac{\varphi_{ext}}{2}\right)} - \frac{1}{2L} \left( \phi^{2} + \xi^{2} \right) \\
-&CdC \dot{\theta} \dot{\xi}  - C_{J}dC_{J} \dot{\theta} \dot{\phi} \\
+& E_{J}dE_{J} \sin{\left(\theta \right)} \sin{\left(\phi - \frac{\varphi_{ext}}{2}\right)} + \frac{dL}{L} \phi \xi
\end{split}
\end{equation}
The  first two lines of Eqn. \eqref{eq:lagnewvars} correspond to the ideal, balanced case of the circuit, that has been the focus of our analysis.
The last two lines, on the other hand, correspond to the deviation from the balanced circuit case. We shall now look at the impact of these deviations when small, i.e. when the adimensional parameters $dC$, $dC_J$, $dE_J$ and $dL$ are all very small numbers. 

To do so, we compute the inverse of the capacitance matrix and retain only first order in these small parameters. Explicitly,
\begin{equation*}
\begin{split}
&\begin{pmatrix}
C+C_{J} & -CdC & - C_{J} dC_{J} \\
 -CdC & C & 0 \\
 - C_{J} dC_{J} & 0 & C_{J}
\end{pmatrix}^{-1}=\\
&\frac{\begin{pmatrix}
C_{J}C & C_{J}CdC & C_{J}C dC_{J} \\
 C_{J}CdC & C_{J}\left( C+C_{J}-C_{J}dC^{2}_{J}\right) & C_{J}C dC_{J}dC \\
  C_{J}C dC_{J} & C_{J}C dC_{J}dC & C \left( C+C_{J}-C dC^{2} \right) 
\end{pmatrix}}{C_{J}C\left( C + C_{J} - C_{J}dC^{2}_{J} - C dC^{2} \right)}\\
&\to \begin{pmatrix}
\frac{1}{C+C_{J} }& \frac{dC}{C+C_{J}} & \frac{ dC_{J}}{C+C_{J}} \\
 \frac{dC}{C+C_{J}} & \frac{1}{C} & 0 \\
  \frac{ dC_{J}}{C+C_{J}}& 0 & \frac{1}{C_{J}}\,.
\end{pmatrix}
\end{split}
\end{equation*}
This approximate inverse capacitance is used in the Legendre transform of Lagrangian \eqref{eq:lagnewvars} to reach the approximate quantum Hamiltonian
\begin{equation}\label{eq:hamapproxcomplete}
\begin{split}
\hat{H}=&4E_{C_J} \hat{Q}^2_{\phi}+4E_{C} \hat{Q}^2_{\xi}+E_L\left(\hat{\phi}^2+\hat{\xi}^{2} \right)\\
+&4E_{C_s} \left( \hat{n}_{\theta} - n_{g} \right)^2 -2E_J\cos\left(\hat{\theta}\right)\cos \left(\hat{\phi}-\frac{\varphi_{\mathrm{ext}}}{2} \right)\\
+&2E_{{C_s}}\left( \hat{n}_{\theta} - n_{g} \right) \left(dC \hat{Q}_{\xi} + dC_{J}\hat{Q}_{\phi}\right) \\
-& E_{J}dE_{J} \sin{\left(\hat{\theta} \right)} \sin{\left(\hat{\phi} - \frac{\varphi_{ext}}{2}\right)} - 2E_{L} dL\hat{\phi} \hat{\xi}\,.
\end{split}
\end{equation}
Here, in parallel with the analysis in the text,
$E_L=\frac{\Phi_0^2}{4\pi^{2}L}$ is the inductive energy, $E_{C_J}=\frac{e^2}{2C_J}$ ($E_{C_s}=\frac{e^2}{2\left(C+C_{J}\right)}$) [$E_{C}=\frac{e^2}{2C}$] denotes the charging energy conjugate to the $\phi$ $(\theta)$  $[\xi]$ mode, and $n_{g}$ is the offset-charge bias due to the electrostatic environment.

We have written this approximate Hamiltonian \eqref{eq:hamapproxcomplete} so that the first two lines give us the  balanced circuit case. Observe that in those two lines there is no coupling of the $\hat{\xi}$ harmonic oscillator to the other modes. These two lines include the case of central interest to use, the symmetric $0$-$\pi$ qubit, when $n_{g}=1/2$ and $\phi_{ext}=\pi$. The last two lines in  Hamiltonian \eqref{eq:hamapproxcomplete}, however, due to deviation from the  balanced circuit situation, explicitly break the symmetries we have considered, because of the coupling to the collective harmonic oscillator  $\hat{\xi}$. In particular, even when  $n_{g}=1/2$ and $\phi_{ext}=\pi$ those terms break explicitly the $D_4$ symmetry we have put to the fore.

Now, as we have mentioned in the main text and made apparent by perturbative computation in section \ref{sec:matrix-elements-sw}, terms that explicitly break the symmetry are exponentially suppressed  in the kinetically dominated regime. Thus, under the energy hierarchy $E_{{C_J}}^{2} \gg E_{{C_J}}  E_L \gg E_J^2 \gg E_{{C_J}} E_{{C_s}}$, for which the perturbative SW analysis is valid and there is kinetic dominance, the breaking of the $D_4$ symmetry by a small lack of balance in the circuit is mild.

One possible objection to this conclusion is that we have not made explicit in the SW analysis what the low energy sector would look like in the presence of the collective oscillator mode $\hat{\xi}$. In other words, we have not included $E_C$ in the energy hierarchy. In order to address this issue, observe that if the full hierarchy holds (kinetic dominance, perturbative SW for the balanced circuit, and semiclassicality of the effective Hamiltonian), then $E_{{C_J}}\gg E_{{C_s}}$, whence $C/C_J+1\gg1$, from which $C\gg C_J$ and $E_{{C_J}}\gg E_{{C}}$.  Next, we observe that $E_{{C_J}}E_L\gg E_{C}E_L\gg E_C E_{{C_s}}$. Thus the only possible issue remaining is the validity of the SW perturbative scheme in the presence of the $\sqrt{E_C E_L}$ frequency of the collective oscillator. Because of the smallness of deviations from balance, that is guaranteed if $E_{{C_s}}= O(E_L)$. As, in fact, in the full hierarchy we have $E_{{C_s}}= o(E_L)$, we complete the analysis and assert that under this ordering of energy scales the deviations from balance are exponentially suppressed.

\end{document}